\documentclass[aps,prd,amsmath,floats,floatfix,twocolumn,superscriptaddress,nofootinbib,noshowpacs]{revtex4-2}

\usepackage{amssymb}
\usepackage{amsmath}
\usepackage{tabularray}
\usepackage{physics}
\usepackage{latexsym}

\usepackage{verbatim}
\usepackage{mathrsfs}
\usepackage{amsfonts}

\usepackage{graphicx,subfigure}
\usepackage{epsfig}

\usepackage{units}
\usepackage{overpic}
\usepackage[utf8]{inputenc}
\usepackage{rotating}
\usepackage{enumerate}

\usepackage{soul}

\usepackage{xfrac}
\usepackage[dvipsnames]{xcolor}
\usepackage{multirow}

\usepackage{booktabs}
\usepackage{varwidth}
\usepackage{caption}
\newsavebox\tmpbox
\newcommand{\Real}{\mathbb{R}}
\newcommand{\Complex}{\mathbb{C}}

\begin{document}

\title{Are nonrelativistic ground state $\ell$-boson stars only stable for $\ell=0$ and $\ell=1$?}

\author{Emmanuel Ch\'avez Nambo}
\affiliation{Instituto de F\'isica y Matem\'aticas,
Universidad Michoacana de San Nicol\'as de Hidalgo,
Edificio C-3, Ciudad Universitaria, 58040 Morelia, Michoac\'an, M\'exico}
\author{Armando A. Roque}
\affiliation{Unidad Acad\'emica de F\'isica, Universidad Aut\'onoma de Zacatecas, 98060, M\'exico.}
\author{Olivier Sarbach}
\affiliation{Instituto de F\'isica y Matem\'aticas,
Universidad Michoacana de San Nicol\'as de Hidalgo,
Edificio C-3, Ciudad Universitaria, 58040 Morelia, Michoac\'an, M\'exico}

\date{\today}

\begin{abstract}
In previous work we analyzed the linear stability of non-relativistic $\ell$-boson stars with respect to radial modes and showed that ground state configurations are stable with respect to these modes, whereas excited states are unstable. In this work we extend the analysis to non-spherical linear mode perturbations. To this purpose, we expand the wave function in terms of tensor spherical harmonics which allows us to decouple the perturbation equations into a family of radial problems. By using a combination of analytic and numerical methods, we show that ground state configurations with $\ell > 1$ possess exponentially in time growing non-radial modes, whereas only oscillating modes are found for $\ell=0$ and $\ell=1$. This leads us to conjecture that nonrelativistic $\ell$-boson stars in their ground state are stable for $\ell=1$ as well as $\ell=0$, while ground state and excited configurations with $\ell > 1$ are unstable.
\end{abstract}

\maketitle

%%%%%%%%%%%%%%%%%%%%%%%%%%%%%%%%%%%%%%%
\section{Introduction}
%%%%%%%%%%%%%%%%%%%%%%%%%%%%%%%%%%%%%%%

Recent investigation has revealed that the multi-field Einstein-Klein-Gordon system admits a rich spectrum of static solutions, even in the spherically symmetric sector~\cite{Alcubierre:2018ahf}. This is due to the fact that when passing from a single to a multitude of $N\geq 3$ scalar fields, the internal symmetry group $U(N)$ can accommodate nontrivial representations of the rotation group $SO(3)$, leading to configurations with nonzero orbital but zero \emph{total} angular momentum, such that they give rise to a spherically symmetric spacetime. For the particular case in which $N = 2\ell+1$ (or an integer multiple thereof) the choice of the irreducible representation with integer spin $\ell$ leads to the $\ell$-boson stars discussed in~\cite{Alcubierre:2018ahf, Alcubierre:2021psa, Jaramillo:2022zwg}, see also~\cite{Olabarrieta:2007di} for an application in the context of critical collapse. In addition to the parameter $\ell$ these configurations are characterized by the node number $n$ of the wave functions' radial profile and a parameter $a_\ell$ controlling their amplitude. Of course, one might object that a theory with an odd number $2\ell+1$ of classical scalar fields is somehow unnatural; however, it was shown that $\ell$-boson stars (and many of their relatives) admit a much more  natural physical interpretation in the realm of semiclassical gravity with a single real scalar (quantum) field~\cite{Alcubierre:2022rgp}.

The stability of $\ell$-boson stars with respect to linear and nonlinear \emph{spherically symmetric} perturbations has been established in~\cite{Alcubierre:2019qnh, Alcubierre:2021mvs} for the ground state configurations (i.e. those with $n=0$ nodes) having $a_\ell$ smaller than the value leading to the maximal mass configuration. Nonetheless, due to their nonzero orbital angular momentum, one cannot expect $\ell$-boson stars with $\ell > 0$ to be stable since they could in principle collapse to a new configuration with zero orbital angular momentum. That such a collapse is, in fact, energetically allowed has been shown in our previous work~\cite{Roque:2023sjl} in the nonrelativistic limit. However, it is clear that such a collapse could only be induced by a nonspherical metric perturbation since otherwise the orbital angular momenta of the scalar fields would be preserved during the time evolution. The stability of $\ell$-boson stars with respect to nonlinear perturbations without symmetries has been studied numerically in~\cite{Guzman:2019gqc, Jaramillo:2020rsv, Sanchis-Gual:2021edp} for the case $\ell=1$, and no instabilities have been found during the timespan of the simulations.

Motivated by these thoughts, in this work, we analyze the stability of $\ell$-boson stars with respect to nonspherical linear perturbations of the fields. To simplify the analysis, we restrict ourselves to the nonrelativistic limit in which these stars are described by stationary solutions of the multi-field Schr\"odinger-Poisson system~\cite{Diosi:1984wuz, Jones:1995yz, Jones:1995wb}. The linear stability property of these Newtonian analogues with respect to spherical perturbations has been studied in our previous work~\cite{Roque:2023sjl}, where it was shown that the ground state configurations are stable with respect to radial perturbations, whereas the excited states with $n > 0$ possess unstable, exponentially in time growing modes. In this article we show that the expectation that $\ell$-boson stars are unstable with respect to nonspherical perturbations even when $n=0$ is correct, at least in the nonrelativistic limit, when $\ell\geq 2$. Interestingly, however, we also find that nonrelativistic $(\ell=1)$-boson stars in their ground state possess only oscillatory modes, and hence they seem to be stable, like the standard nonrelativistic boson stars with $\ell=0$~\cite{Lieb1977, Cazenave1982}.

We mention in passing that the nonrelativistic approximation our results are limited to contains one of the most relevant physical potential applications of the $\ell$-boson stars; namely the modeling of galactic dark matter halo cores in the context of ultralight scalar field dark matter, see for instance Refs.~\cite{Schive:2014hza,Marsh:2015wka,Schwabeetal2016,Hui:2016ltb, Gonzalez-Morales:2016yaf, LinaresCedeno:2020dte, Goldstein:2022pxu,Navarro-Boullosa:2023bya} for recent progress.

The remaining of this work is organized as follows. In Sec.~\ref{Sec:SP} we provide a brief review of the $N$-particle Schr\"odinger-Poisson system, the associated energy functional which will play an important role in our stability analysis, and the stationary solutions describing the nonrelativistic $\ell$-boson stars. Next, in Sec.~\ref{Sec:LinSys} we derive the mode equation describing linear perturbations oscillating in time with a complex frequency $\lambda$, and we show how to decouple it by expanding the fields in terms of vector spherical harmonics (for $\ell=1$) or tensor spherical harmonics (for $\ell>1$). This leads to a decoupled family of radial eigenvalue problems with eigenvalue $\lambda$, where each of these problems is labeled by the value of the total angular momentum $J$, its associated magnetic quantum number $M$ and a parity flag. In Sec.~\ref{Sec:Properties} we discuss some important properties of these problems; in particular, we show that they admit stationary modes with $J\neq 0$, and we prove that no instabilities can arise in the odd-parity sector nor in the even-parity sector with high enough values of $J$. Our numerical results are presented in Sec.~\ref{Sec:Numerical} where we solve the eigenvalue problems using a spectral method similar to our previous work~\cite{Roque:2023sjl}. Conclusions are drawn in Sec.~\ref{Sec:Conclusions} and technical results are further developed in appendices~\ref{App:SphericalTensorHarmonics}--\ref{App:FirstOrderFormulation}.

%%%%%%%%%%%%%%%%%%%%%%%%%%%%%%%%%%%%%%%
\section{The $N$-particle Schr\"odinger-Poisson system}
\label{Sec:SP}
%%%%%%%%%%%%%%%%%%%%%%%%%%%%%%%%%%%%%%%

Consider a nonrelativistic system of $N$ spinless, indistinguishable and uncorrelated particles of mass $\mu$ whose only interaction is through the gravitational potential $\mathbf{U}$ generated by them. Specifically, we consider an orthonormal set of wave functions $\phi_j$ in the one-particle Hilbert space $L^2(\Real^3)$ such that $(\phi_j, \phi_k) = \delta_{jk}$. Assuming that there are $N_j$ particles in the state $\phi_j$, the wave functions $\phi_j$ satisfy the $N$-particle Schr\"odinger-Poisson system 
\begin{subequations}\label{SP}
\begin{align}
i\hbar \frac{\partial \phi_j (t, \vec{x})}{\partial t} &= \left[-\frac{\hbar^2}{2\mu} \triangle + \mu \mathbf{U}(t, \vec{x})\right]\phi_j(t, \vec{x}),\\
\triangle \mathbf{U}(t, \vec{x}) &= 4\pi G\mu \sum_{j} N_j |\phi_j(t, \vec{x})|^2,
\end{align}
\end{subequations}
where $\sum_j N_j = N$ is the total number of particles. The evolution described by the Schr\"odinger-Poisson system is unitary, i.e., the $L^2$-norms of the wave functions $\phi_j$ are preserved. Further, the evolution preserves each scalar product $(\phi_j, \phi_k)$, such that it is sufficient to impose the condition $(\phi_j, \phi_k) = \delta_{jk}$ at the initial time $t = 0$. Additionally, it can be verified that the functional 

\begin{align}
\label{EnergyFunc}
\mathcal{E}&[u] = \frac{\hbar^2}{2\mu} \sum_j N_j \int \abs{\nabla u_j(\vec{x})}^2 d^3 x\nonumber\\
    &-\frac{G\mu^2}{2}\sum_{j,k} N_j N_k \int\int\frac{|u_j(\vec{x})|^2 |u_k(\vec{y})|^2}{|x-y|} d^3x d^3y,
\end{align}
is conserved in time, that is $\mathcal{E}[\phi_j(t)]$ is independent of $t$ for any solution $\phi_j(t, \vec{x})$ of the system (\ref{SP}) for which $|\mathcal{E}[\phi_j(t)]| < \infty $. As in Ref.~\cite{Roque:2023sjl} its second variation will be very useful to study the stability properties of $\ell$-boson stars.

Before continuing, it is convenient to rewrite the system in terms of dimensionless quantities as,
\begin{subequations}
\label{SPDimensionless}
\begin{align}
i\frac{\partial \bar{\phi}_j}{\partial \bar{t}} (\bar{t}, \vec{\bar{x}}) &= \left[- \bar{\triangle} + \bar{U}(\bar{t}, \vec{\bar{x}})\right]\bar{\phi}_j(\bar{t}, \vec{\bar{x}}),\\
\bar{\triangle} \bar{U}(\bar{t}, \vec{\bar{x}}) &= \sum_j N_j |\bar{\phi}_j(\bar{t}, \vec{\bar{x}})|^2,
\end{align}
\end{subequations}
%
%\begin{subequations}\label{SPDimensionless}
%\begin{align}
%i\frac{\partial \phi_j}{\partial t} (t, \vec{x}) &= \left[- \triangle + U(t, \vec{x})\right]\phi_j(t, \vec{x}),\\
%\triangle U(t, \vec{x}) &= \sum_{j = 1}^{J} N_j|\phi_j(t, \vec{x})|^2.
%\end{align}
%\end{subequations}
where we used the transformations
\begin{align}\label{AdVar}
    \begin{array}{ll}
        t=t_c \bar{t}/\Lambda^2, & \vec{x} = d_c\vec{\bar{x}}/\Lambda,\\[0.3cm]
        \phi_j = \Lambda^2 \bar{\phi}_{j}/\sqrt{4\pi d_c^3}, & \mathbf{U}=2 \Lambda^2 v_c^2 \bar{U},
    \end{array}
\end{align}	
with $\Lambda$ an arbitrary positive dimensionless scale factor, $v_c := d_c/t_c$ a characteristic velocity defined in terms of the characteristic distance and length 
\begin{align}\label{Eq:characteristicQ}
d_c := \frac{\hbar^2}{2G\mu^3},\quad t_c := \frac{\hbar^3}{2G^2\mu^5}.
\end{align}
In order to simplify the notation, in what follows we shall omit the bars and denote dimensionfull quantities with the  superscript $phys$ whenever necessary. Furthermore, we introduce the following notation
\begin{align}\label{PsiDefin}
\Psi := (\psi_1, \ldots, \psi_{j_{max}})^T,\quad |\Psi|^2 := \sum_{j=1}^{j_{max}} |\psi_j|^2,
\end{align}
where $j_{max}$ denotes the maximum number of different excited states in the configuration, the superscript $T$ refers to the transposed and $\psi_j := \sqrt{N_j}\phi_j$. With this we rewrite the system~(\ref{SPDimensionless}) as follows
\begin{subequations}\label{SPVec}
\begin{align}
\label{SPVecA}
i\frac{\partial \Psi(t, \vec{x})}{\partial t}&= \left[- \triangle + U(t, \vec{x})\right]\Psi(t, \vec{x}),\\
\label{SPVecB}
\triangle U(t, \vec{x}) &= |\Psi(t, \vec{x})|^2,
\end{align}
\end{subequations}
with the condition
\begin{align}
(\psi_j, \psi_k) = \frac{4\pi}{\Lambda}\sqrt{N_j N_k}\delta_{jk}.
\label{Eq:Ortho}
\end{align}
Equivalently, the system~(\ref{SPVec}) can be written as a single nonlinear equation  
\begin{align}
\label{PC}
i\frac{\partial \Psi}{\partial t} (t, \vec{x}) = \hat{\mathcal{H}} \Psi(t, \vec{x}),
\end{align}
with the integro-differential operator

\begin{equation}
\hat{\mathcal{H}} := -\triangle + \triangle^{-1} (|\Psi|^2),
\end{equation}
where $\triangle^{-1}$ denotes the inverse operator of $\triangle$, defined by
\begin{align}
\triangle^{-1} (A)(\vec{x}) = -\frac{1}{4\pi} \int \frac{A(\vec{y})}{|\vec{x} - \vec{y}|} d^3y,
\label{Eq:InverseLaplace}
\end{align}
when acting on an arbitrary function $A$.  

The conserved energy functional~(\ref{EnergyFunc}) in terms of the dimensionless quantities defined in Eqs.~(\ref{AdVar},~\ref{PsiDefin}) takes the form $\mathcal{E}^{phys}[u]= \mu v_c^2 \Lambda^3\mathcal{E}[u]/\pi$, where
\begin{equation}
\mathcal{E}[u] =\frac{1}{2}\int \abs{\nabla u(\vec{x})}^2 d^3x 
- D[n,n],\qquad n := |u|^2,
%\nonumber\\
%&-\frac{1}{16\pi}\int\int\frac{|u(\vec{x})|^2 |u(\vec{y})|^2}{|\vec{x}-\vec{y}|} d^3x d^3y.
\end{equation}
with the bilinear functional $D[n,n]$ defined by
\begin{equation}
\label{DOperator}
D[n,n] := \frac{1}{16\pi}\int\int\frac{n(\vec{x}) n(\vec{y})}{|\vec{x}-\vec{y}|} d^3x d^3y.
\end{equation}

For the following, the first and second variations of $\mathcal{E}$ will be useful:
\begin{subequations}
\begin{align}
\delta\mathcal{E} &= \Re( \hat{\mathcal{H}}u,\delta u),
\label{Eq:FirstVariation}\\
\delta^2\mathcal{E} &= \Re(\hat{\mathcal{H}} u,\delta^2 u)
 + (\delta u,\hat{\mathcal{H}}\delta u)
 - 2D[\delta n,\delta n],
% \nonumber\\
% &- \frac{1}{2\pi}\int\int\frac{
% \Re[u(\vec{x})^* \delta u(\vec{x})]
% \Re[u(\vec{y})^* \delta u(\vec{y})] }{|\vec{x}-\vec{y}|} d^3x d^3y,
\label{Eq:SecondVariation}
\end{align}
\end{subequations}
with $\delta n := 2\Re(u^*\delta u)$ and $(u,v)$ denoting the standard $L^2$-scalar product between $u = (u_1,\ldots,u_{j_{max}})$ and $v = (v_1,\ldots,v_{j_{max}})$, that is
\begin{equation}
(u,v) := \sum\limits_{j=1}^{{j_{max}}} (u_j,v_j)
 = \sum\limits_{j=1}^{{j_{max}}}\int u_j(\vec{x})^* v_j(\vec{x}) d^3 x.
\label{Eq:L2Product}
\end{equation}

%%%%%%%%%
\subsection{The stationary equations} 
%%%%%%%%

Stationary solutions are characterized by a harmonic dependency on time, such that
\begin{equation}
\Psi(t,\vec{x}) = e^{-iEt}\chi_0(\vec{x}), \quad \vec{x}\in \Real^3,
\label{Eq:StationaryAnsatz}
\end{equation}
with $\chi_0$ a column vector where each component is a complex-valued function and $E = \mbox{diag}(E_1, E_2, \ldots, E_{j_{max}})$ is a real diagonal matrix. For $\ell$-boson stars, all $E_j$ are equal to each other. However, other solutions including multi-$\ell$ multistate solutions~\cite{Alcubierre:2022rgp} have different $E_j$'s. $(E, \chi_0)$ are determined by the non-linear (multi-)eigenvalue problem
\begin{equation}
\label{StationaryProblem}
    \hat{\mathcal{H}}_0 \chi_0 = E\chi_0,
\end{equation}
with 
\begin{equation}
\hat{\mathcal{H}}_0 := -\triangle + \triangle^{-1}(|\chi_0|^2).
\end{equation}
Taking into account the orthonormality conditions~(\ref{Eq:Ortho}), the first and second variations of the energy functional~(\ref{Eq:FirstVariation}, \ref{Eq:SecondVariation}) associated with the background field $\chi_0$ yield
\begin{subequations}
\begin{align}
\delta\mathcal{E} &= 
\frac{2\pi}{\Lambda}\sum\limits_j E_j \delta N_j,\\
\delta^2\mathcal{E} &= \frac{2\pi}{\Lambda}\sum\limits_j E_j \delta^2 N_j
 + (\delta u,[\hat{\mathcal{H}}_0 - E]\delta u) - 2D[\delta n,\delta n],
\label{Eq:SecondVariationchi0}
%\nonumber\\
% &- \frac{1}{2\pi}\int\int\frac{
% \Re[\chi_0(\vec{x})^* \delta u(\vec{x})]
% \Re[\chi_0(\vec{y})^* \delta u(\vec{y})] }{|\vec{x}-\vec{y}|} d^3x d^3y.
\end{align}
\end{subequations}
with $\delta n := 2\Re(\chi_0^*\delta u)$. In particular, if the particle numbers $N_j$ are held fixed, it follows that $\chi_0$ is a critical point of the energy function $\mathcal{E}$ and the second variation is expected to give information on the stability of the stationary solution. Note that $D[\delta n,\delta n]$ is positive definite.

\subsection{Nonrelativistic $\ell$-boson stars}

Particular stationary solutions consist of non-relativistic $\ell$-boson stars~\cite{nambo21, nambo19, jaramillo19, Roque:2023sjl}. Fixing some value $\ell\in \{ 0,1,2,\ldots \}$, they are obtained from the ansatz
\begin{equation}
\chi_0(\vec{x}) = \sigma_\ell^{(0)}(r)\mathcal{Y}_\ell(\vartheta,\varphi),
\label{Eq:chi0lboson}
\end{equation}
where the function $\sigma_\ell^{(0)}$ is real-valued and where
\begin{equation}
\mathcal{Y}_\ell := \sqrt{\frac{4\pi}{2\ell+1}}\left( Y^{\ell,-\ell},Y^{\ell,-\ell+1},\ldots,Y^{\ell,\ell} \right)^T,
\end{equation}
with $Y^{\ell m}$ denoting the standard spherical harmonics. Since $|\mathcal{Y}_\ell|^2 = 1$, it follows that $|\Psi|^2 = |\sigma_\ell^{(0)}|^2$ and Eq.~(\ref{StationaryProblem}) reduces to Eq.~(20) in~\cite{Roque:2023sjl} under the assumption that the matrix $E$ is equal to $E_\ell$ times the identity matrix. The orthonormality condition~(\ref{Eq:Ortho}) reduces to
\begin{align}
\int\limits_0^\infty |\sigma_\ell^{(0)}(r)|^2 r^2 dr = \frac{(2\ell+1) K}{\Lambda},
\end{align}
with $K = N_j$ the equal number of particles in each state. For convenience in this paper we set the scale factor to $\Lambda:= N = (2\ell+1)K$.

%%%%%%%%%%%%%%%%%%%%%%%%%%%%%%%%%%%%%%%
\section{The linearized system}
\label{Sec:LinSys}
%%%%%%%%%%%%%%%%%%%%%%%%%%%%%%%%%%%%%%%

In this section we linearize the system~(\ref{SPVec}) or~(\ref{PC}) around a stationary background solution. In Sec.~\ref{Sec:LinSysGeneral} we discuss the most general case, which is valid for arbitrary stationary backgrounds, and we derive the equations describing linear modes. In Sec.~\ref{Sec:LinSysRadial} we show that for the particular case of purely radial perturbations of $\ell$-boson stars this system reduces to the one of our previous work. Next, in Sec.~\ref{Sec:LinSysl1} we discuss the mode equations for the $(\ell=1)$-boson stars and show that they can be decoupled using spherically vector harmonics. This construction is then generalized to boson stars with arbitrary $\ell$ in Sec.~\ref{Sec:LinSyslbosonstars}.

\subsection{Derivation of the mode equations}
\label{Sec:LinSysGeneral}

In order to linearize Eq.~(\ref{PC}) about a stationary solution $\chi_0$, we assume an expansion of $\Psi$ in terms of a small parameter $\epsilon > 0$ of the form
\begin{equation}
\label{Eq:Expansion}
    \Psi(t,\vec{x}) = e^{-iEt}[\chi_0(\vec{x}) + \epsilon\chi(t, \vec{x}) + \mathcal{O}(\epsilon^2)].
\end{equation}
Here, $\chi$ is a column vector in which each component is a complex-valued function and ($E$, $\chi_0$) is a solution to the problem~(\ref{StationaryProblem}).

Substituting the expansion~(\ref{Eq:Expansion}) into Eq.~(\ref{PC}) and considering the first-order terms we arrive at the perturbed evolution equation
\begin{align}    
i\frac{\partial\chi}{\partial t} = \bigg(\hat{\mathcal{H}}_0-E\bigg)\chi + 2\triangle^{-1}\bigg(\Re{\chi_0^{*}\chi}\bigg)\chi_0,
\label{Eq:PerturbationEq}
\end{align}
where $\chi_0^*$ denotes the transposed conjugate of $\chi_0$.

Following Refs.~\cite{2002math.ph...8045H, Roque:2023sjl} we separate the time and spatial parts of $\chi$ using the ansatz
\begin{align}
\label{Eq:SigmaAnsatz}
\chi(t,\vec{x}) = e^{\lambda t}\left[ \mathcal{A}(\vec{x})+\mathcal{B}(\vec{x})\right] + e^{\lambda^* t}\overline{\left[
\mathcal{A}(\vec{x}) - \mathcal{B}(\vec{x}) \right]},
\end{align}
where the bar denotes complex conjugation. Here $\mathcal{A}$ and $\mathcal{B}$ are complex vector-valued functions depending only on $\vec{x}$ and $\lambda$ is a complex number. Note that when $\lambda = \lambda^*$ is real, one can assume that $\mathcal{A}$ is real and $\mathcal{B}$ is purely imaginary.

Introducing Eq.~(\ref{Eq:SigmaAnsatz}) into Eq.~(\ref{Eq:PerturbationEq}) one obtains, after setting the coefficients in front of $e^{\lambda^* t}$ and $e^{\lambda t}$ to zero,
\begin{subequations}
\label{EquationAB}
 	\begin{eqnarray}
        i\lambda \mathcal{A} &=&\left(\hat{\mathcal{H}}_0 - E\right)\mathcal{B} \label{EquationA}\\
        &+& i\left\{\triangle^{-1}[\chi_0^*(\mathcal{A} + \mathcal{B}) + \chi_0^T(\mathcal{A} - \mathcal{B})] \right\}\Im{\chi_0},
\nonumber\\ 
 		i\lambda \mathcal{B} &=&\left(\hat{\mathcal{H}}_0 - E\right)\mathcal{A} \label{EquationB}\\
        &+& \left\{\triangle^{-1}[\chi_0^*(\mathcal{A} + \mathcal{B}) + \chi_0^T(\mathcal{A} - \mathcal{B})]\right\}\Re{\chi_0}.
\nonumber
 	\end{eqnarray}
 \end{subequations}
These two equations remain correct for the case in which $\lambda$ is real, provided $\mathcal{A} = \mathcal{A}_R$ is assumed to be real and $\mathcal{B} = i \mathcal{B}_I$ is purely imaginary. In this case,
\begin{align}
\chi_0^*(\mathcal{A} + \mathcal{B}) + \chi_0^T(\mathcal{A} - \mathcal{B})
 &= \\
 &2(\Re\chi_0)^T\mathcal{A}_R + 2(\Im\chi_0)^T\mathcal{B}_I,\nonumber
\end{align}
which is real. Note also that when $\chi_0$ is real, Eqs.~(\ref{EquationAB}) simplify considerably.

Finally, we recall the orthogonality condition~(\ref{Eq:Ortho}), which yields
\begin{equation}
(\chi_{0,j},\chi_k) + (\chi_j,\chi_{0,k}) = \frac{4\pi}{\Lambda}\delta_{jk}\delta N_k,
\label{Eq:OrthoBis}
\end{equation}
with $\delta N_k$ denoting the first variation of $N_k$. Using the ansatz~(\ref{Eq:SigmaAnsatz}) and assuming, for simplicity, that $\chi_0$ is real, this implies
\begin{subequations}
\begin{align}
(\chi_{0,j},\mathcal{A}_k) + (\overline{\mathcal{A}_j},\chi_{0,k}) &= 0,
\\
(\chi_{0,j},\mathcal{B}_k) - (\overline{\mathcal{B}_j},\chi_{0,k}) &= 0,
\end{align}
\end{subequations}
One can easily verify that these conditions are a consequence of Eqs.~(\ref{EquationAB}) when $\lambda\neq 0$.

\subsection{Example: radial perturbations of Newtonian $\ell$-boson stars}
\label{Sec:LinSysRadial}

For linear perturbations which keep the angular dependency fixed, the relation between~(\ref{Eq:SigmaAnsatz}) and the corresponding ansatz~(25) in~\cite{Roque:2023sjl} is given by
\begin{equation}
\mathcal{A} + \mathcal{B} = (A+B)\mathcal{Y}_\ell,\qquad
\overline{\mathcal{A} - \mathcal{B}} = \overline{(A-B)}\mathcal{Y}_\ell,
\end{equation}
or, equivalently,
\begin{subequations}
\begin{align}
\mathcal{A} &= A\Re(\mathcal{Y}_\ell) + i B\Im(\mathcal{Y}_\ell),\\
\mathcal{B} &= B\Re(\mathcal{Y}_\ell) + i A\Im(\mathcal{Y}_\ell).
\end{align}
\end{subequations}
Using the fact that for $\ell>0$ the vector-valued functions $\Re(\mathcal{Y}_\ell)$ and $\Im(\mathcal{Y}_\ell)$ are linearly independent from each other, it is not difficult to verify that this ansatz reduces Eqs.~(\ref{EquationAB}) to the system~(26) in~\cite{Roque:2023sjl}.

\subsection{Example: linear perturbations of $(\ell=1)$-boson stars using vector spherical harmonics}
\label{Sec:LinSysl1}

An alternative representation of $\ell$-boson stars which is more convenient for the perturbation analysis that follows can be given in terms of tensor spherical harmonics. We first illustrate this technique for Newtonian $\ell$-boson stars with $\ell=1$ and discuss the generalization to $\ell > 1$ in the next subsection. For this, we start by noticing that
\begin{equation}
\mathcal{Y}_1(\vartheta,\varphi) = \left( \begin{array}{c}
\frac{1}{\sqrt{2}}(\hat{x} - i\hat{y}) \\
\hat{z}\\
-\frac{1}{\sqrt{2}}(\hat{x} + i\hat{y})
\end{array} \right)
 = U\hat{\vec x}
\end{equation}
where $\hat{\vec x} = (\hat{x},\hat{y},\hat{z}) := (\cos\varphi\sin\vartheta,\sin\varphi\sin\vartheta,\cos\vartheta)$ and $U$ is the unitary matrix
\begin{equation}
U := \frac{1}{\sqrt{2}}\left( \begin{array}{rrr}
 1 & -i & 0 \\
 0 & 0 & \sqrt{2} \\
 -1 & -i & 0
 \end{array}\right).
\end{equation}
Hence, for $1$-boson stars, we may replace $\mathcal{Y}_1(\vartheta,\varphi)$ in the right-hand side of Eq.~(\ref{Eq:chi0lboson}) with $\hat{\vec x}$. A generic linear perturbation of such stars can then be described by expanding the fields $\mathcal{A}$ and $\mathcal{B}$ in terms of vector spherical harmonics, which are defined by~\cite{Khersonskii:1988krb}
\begin{subequations}\label{Eq:TensH}
\begin{align}
\vec{Y}^{JM}(\vartheta,\varphi) &:= \hat{\vec x} Y^{JM}(\vartheta,\varphi),\\
\vec{\Psi}^{JM}(\vartheta,\varphi) &:=
\frac{1}{\sqrt{J(J+1)}} r\vec{\nabla} Y^{JM}(\vartheta,\varphi),\\
\vec{\Phi}^{JM}(\vartheta,\varphi) &:= \frac{1}{i\sqrt{J(J+1)}} \vec{x}\wedge\vec{\nabla} Y^{JM}(\vartheta,\varphi),
\end{align}
\end{subequations}
where $r := |\vec{x}|$ and $J$ refers to the total angular momentum number and $M$ to the corresponding magnetic quantum number. Using the identities $\partial_k r = \hat{x}_k$ and $\partial_j\hat{x}_k = (\delta_{jk} - \hat{x}_j\hat{x}_k)/r$ and observing that $\vec{\Phi}^{JM}$ is proportional to the orbital angular momentum operator acting on $Y^{JM}$, it is not difficult to verify that
\begin{subequations}
\begin{align}
\Delta \vec{Y}^{JM} &= -\frac{J(J + 1) + 2}{r^2}\vec{Y}^{JM} + \frac{2\sqrt{J(J+1)}}{r^2}\vec{\Psi}^{JM},\\
\Delta \vec{\Psi}^{JM} &= \frac{2\sqrt{J(J+1)}}{r^2}\vec{Y}^{JM} -\frac{J(J + 1)}{r^2}\vec{\Psi}^{JM},\\
\Delta \vec{\Phi}^{JM} &= -\frac{J(J + 1) }{r^2}\vec{\Phi}^{JM} .
\end{align}
\end{subequations}
Note that $\vec{\Psi}^{JM}$ and $\vec{\Phi}^{JM}$ are orthogonal to $\vec{x}$ and that they vanish for $J=0$.

Expanding
\begin{equation}
\mathcal{A} = \sum\limits_{JM}\left( A_{JM}^r\vec{Y}^{JM}
+ A_{JM}^{(1)}\vec{\Psi}^{JM}
+ A_{JM}^{(2)}\vec{\Phi}^{JM} \right)
\end{equation}
with complex-valued functions $A_{JM}^r$, $A_{JM}^{(1)}$ and $A_{JM}^{(2)}$ depending on $r$ and similarly for $\mathcal{B}$ a simple calculation first reveals that
\begin{equation}
\chi_0^*(\mathcal{A} + \mathcal{B}) + \chi_0^T(\mathcal{A} - \mathcal{B}) = 2\sigma_1^{(0)}\sum\limits_{JM} A_{JM}^r Y^{JM},
\end{equation}
from which
\begin{align}
\Delta^{-1}&\big[\chi_{0}^{*}(\mathcal{A}+\mathcal{B})\nonumber\\
& + \chi_0^T(\mathcal{A} - \mathcal{B}) \big]= 2\sum\limits_{JM}\triangle_J^{-1}\left(\sigma_1^{(0)}A_{JM}^r \right) Y^{JM},
\end{align}
with $\Delta_J^{-1}$ denoting the inverse of the operator
\begin{equation}
\triangle_J := \frac{1}{r^2}\frac{d}{dr}\left( r^2\frac{d}{dr} \right) - \frac{J(J+1)}{r^2}.\label{Eq:DeltaJ}
\end{equation}
From Eq.~(\ref{Eq:InverseLaplace}) and the well-known decomposition of $1/|\vec{x}-\vec{y}|$ in terms of spherical harmonics one obtains the explicit representation
\begin{equation}
\triangle_J^{-1}(f)(r) = -\frac{1}{2J+1}\int\limits_0^\infty \frac{r_<^J}{r_>^{J+1}} f(\tilde{r})\tilde{r}^2 d\tilde{r},
\label{Eq:LapJInv}
\end{equation}
with $r_<:=\min\{ r,\tilde{r} \}$ and $r_>:=\max\{ r,\tilde{r} \}$.

\begin{widetext}
Using this, Eqs.~(\ref{EquationAB}) yields the following system of equations:
\begin{subequations}\label{Eq:LinSyst}
\begin{align}
i\lambda \left( \begin{array}{c}
A_{JM}^r \\ A_{JM}^{(1)} \end{array} \right) &= (\mathcal{\hat H}_J^{(0)} - E)\left( \begin{array}{c}
B_{JM}^r \\ B_{JM}^{(1)} \end{array} \right) + \frac{2}{r^2}\left( \begin{array}{cc}
 1 & -\sqrt{J(J+1)} \\
 -\sqrt{J(J+1)} & 0
\end{array} \right)\left( \begin{array}{c}
B_{JM}^r \\ B_{JM}^{(1)} \end{array} \right),
\label{Eq:LinSystA}\\
i\lambda \left( \begin{array}{c}
B_{JM}^r \\ B_{JM}^{(1)} \end{array} \right) &= (\mathcal{\hat H}_J^{(0)} - E)\left( \begin{array}{c}
A_{JM}^r \\ A_{JM}^{(1)} \end{array} \right) + \frac{2}{r^2}\left( \begin{array}{cc}
 1 & -\sqrt{J(J+1)} \\
 -\sqrt{J(J+1)} & 0
\end{array} \right)\left( \begin{array}{c}
A_{JM}^r \\ A_{LJ}^{(1)} \end{array} \right)
 + 2\sigma_1^{(0)}\Delta_J^{-1}\left ( \sigma_1^{(0)}A_{JM}^r \right)
 \left( \begin{array}{c} 1 \\ 0 \end{array} \right),
\label{Eq:LinSystB}\\
i\lambda \left( \begin{array}{c}
A_{JM}^{(2)} \\ B_{JM}^{(2)} \end{array} \right) &= (\mathcal{\hat H}_J^{(0)} - E)\left( \begin{array}{c}
B_{JM}^{(2)} \\ A_{JM}^{(2)} \end{array} \right),
\label{Eq:LinSyst2}
\end{align}
\end{subequations}
\end{widetext}
where $\mathcal{\hat H}_J^{(0)} := -\Delta_J + \Delta^{-1}\left( |\sigma_1^{(0)}|^2 \right)$. When $J=0$, $A_{JM}^{(1,2)}$ and $B_{JM}^{(1,2)}$ are void, and the system reduces to the same system as Eq.~(26) in~\cite{Roque:2023sjl} with $\ell=1$.

One can simplify the operators on the right-hand side by diagonalizing the $2\times 2$ symmetric matrix
\begin{equation}
\left( \begin{array}{cc}
 1 & -\sqrt{J(J+1)} \\
 -\sqrt{J(J+1)} & 0
\end{array} \right)
 = T D T^{-1}
\end{equation}
with $D = \mbox{diag}(-J,J+1)$ and the orthogonal matrix
\begin{equation}
T = \frac{1}{\sqrt{2J+1}}\left( \begin{array}{cc}
 \sqrt{J} & -\sqrt{J+1} \\
 \sqrt{J+1} & \sqrt{J}
\end{array} \right).
\end{equation}
This allows one to rewrite Eqs.~(\ref{Eq:LinSystA}, \ref{Eq:LinSystB}) as
\begin{widetext}
\begin{subequations}\label{Eqell1}
\begin{align}
i\lambda\alpha_{JM} &= 
\left( \begin{array}{cc}
\mathcal{\hat H}_{J-1}^{(0)} - E & 0 \\
0 & \mathcal{\hat H}_{J+1}^{(0)} - E
\end{array} \right)\beta_{JM},
\label{Eq:AlphaJM}\\
i\lambda\beta_{JM} &= 
\left( \begin{array}{cc}
\mathcal{\hat H}_{J-1}^{(0)} - E & 0 \\
0 & \mathcal{\hat H}_{J+1}^{(0)} - E
\end{array} \right)\alpha_{JM}
 + \frac{2\sigma_1^{(0)}}{2J+1}
\left( \begin{array}{cc} J & -\sqrt{J(J+1)} \\ -\sqrt{J(J+1)} & J+1 \end{array} \right)
 \Delta_J^{-1}\left ( \sigma_1^{(0)}\alpha_{JM} \right),
 \label{Eq:BetaJM}
\end{align}
\end{subequations}
\end{widetext}
where $\alpha_{JM} := T^{-1}(A_{JM}^r,A_{JM}^{(1)})^T$ and $\beta_{JM} := T^{-1}(B_{JM}^r,B_{JM}^{(1)})^T$. When $J=0$, the first components of $\alpha_{JM}$ and $\beta_{JM}$ are void and only the second components of Eqs.~(\ref{Eq:AlphaJM}) and (\ref{Eq:BetaJM}) should be considered.

\subsection{Linear perturbation for arbitrary $\ell$ using tensor spherical harmonics}
\label{Sec:LinSyslbosonstars}

For the general case we expand the fields in terms of tensor spherical harmonics $Y^{JM}{}_{L\ell}$ which are eigenfunctions of the operators $\hat{J}^2$, $\hat{L}^2$, $\hat{S}^2$ and $\hat{J}_z$~\cite{Khersonskii:1988krb}. They are defined by
\begin{equation}
\label{TensorSphericalHar}
Y^{JM}{}_{L\ell}(\vartheta,\varphi) := \sum\limits_{m,\sigma}
C^{JM}{}_{Lm\ell\sigma} Y^{Lm}(\vartheta,\varphi)\xi^{\ell\sigma},
\end{equation}
with $C^{JM}{}_{Lm\ell\sigma}$ the Clebsch-Gordan coefficients and $\xi^{\ell\sigma}$ denoting an orthonormal basis of spin functions in $\Complex^{2\ell+1}$, see Appendix~\ref{App:SphericalTensorHarmonics} for more details. Note that
\begin{equation}
Y^{00}{}_{\ell\ell} = \frac{(-1)^\ell}{\sqrt{2\ell+1}}\sum\limits_\sigma (Y^{\ell\sigma})^*\xi^{\ell\sigma},
\end{equation}
and for a suitable choice of the basis functions $\xi^{\ell\sigma}$ and using Eq.~(\ref{JZeroCB}) in appendix \ref{App:SphericalTensorHarmonics} one obtains
\begin{equation}
Y^{00}{}_{\ell\ell} = \frac{1}{\sqrt{4\pi}}\mathcal{Y}_\ell.
\end{equation}

However, for the following we shall assume that the basis spin functions satisfy the relation
\begin{equation}
\label{ComplexConjugateBasisSpin}
\overline{\xi^{\ell\sigma}} = (-1)^\sigma\xi^{\ell -\sigma}
\end{equation}
for all $\sigma = -\ell,\ldots,\ell$, which implies that
\begin{equation}
\overline{Y^{JM}{}_{L\ell}} = (-1)^{J+M+L+\ell} Y^{J -M}{}_{L\ell},
\label{ComplexConjugationTensorSpherical}
\end{equation}
and, in particular, that $Y^{00}{}_{\ell\ell}$ is real-valued. For $\ell=1$, for instance, this basis can be chosen as
\begin{equation}
\label{BasisSpin1}
\xi^{1-1} := \frac{1}{\sqrt{2}} (\hat{e}_x - i \hat{e}_y),\quad
\xi^{11} := -\frac{1}{\sqrt{2}} (\hat{e}_x + i \hat{e}_y),
\end{equation}
and $\xi^{10} := \hat{e}_z$, with $\hat{e}_x,\hat{e}_y,\hat{e}_z$ the usual Cartesian basis of $\Complex^3$, and this yields $Y^{00}{}_{11} = -\hat{\vec{x}}/\sqrt{4\pi}$ which, up to the normalization factor $-1/\sqrt{4\pi}$, agrees with the choice in the previous subsection. 

Due to their completeness, the  tensor spherical harmonics can be used to expand the fields $\mathcal{A}$ and $\mathcal{B}$ as follows:
\begin{equation}
\mathcal{A} = \sum\limits_{JLM} A_{JM}{}^L(r) Y^{JM}{}_{L\ell},
\label{Eq:Descomp}
\end{equation}
and similarly for $\mathcal{B}$. The fact that the background has zero total angular momentum implies that the different $JM$ modes decouple in the linearized equations. To derive the mode equations and exhibit this decoupling, we use the identity
\begin{equation}
(Y^{00}{}_{\ell\ell})^* Y^{JM}{}_{L\ell}
 = \frac{(-1)^\ell}{\sqrt{4\pi}}\sqrt{\frac{2L+1}{2J+1}}
 C^{J0}{}_{L0\ell0} Y^{JM},
\end{equation}
which can be deduced from the product formula for the spherical harmonics, see for instance Eq.~(10) in Sec.~5.6 in Ref.~\cite{Khersonskii:1988krb}. One obtains from this
\begin{equation}
\Delta^{-1}(\chi_0^*\mathcal{A})\chi_0
 = \sum\limits_{JLM} Q_{JM}{}^L(r) Y^{JM}{}_{L\ell},
\label{Eq:Q}
\end{equation}
with
\begin{eqnarray}
Q_{JM}{}^L(r) &=& \sigma_\ell^{(0)}(r)\sum\limits_{L'=|J-\ell|}^{J+\ell}
\frac{\sqrt{(2L+1)(2L'+1)}}{2J+1} 
\nonumber\\
 &\times& C^{J0}{}_{L0\ell 0} C^{J0}{}_{L'0\ell 0} 
 \Delta_J^{-1}\left( \sigma_\ell^{(0)} A_{JM}{}^{L'} \right)(r).
\qquad
\label{Eq:QJML}
\end{eqnarray}
The selection rules for the Clebsch-Gordan coefficients imply that $C^{J0}{}_{L0\ell 0}$ is different from zero only if $|J-\ell|\leq L\leq J+\ell$ and $J+L+\ell$ is even. Therefore, the only non-vanishing coefficients are $Q_{JM}{}^{|J-\ell|}, Q_{JM}{}^{|J-\ell|+2},\ldots,Q_{JM}{}^{J+\ell}$. Likewise, only the amplitudes $A_{JM}{}^{|J-\ell|}, A_{JM}{}^{|J-\ell|+2},\ldots,A_{JM}{}^{J+\ell}$ appear in the sum in the right-hand side of Eq.~(\ref{Eq:QJML}). 

Using this observation, Eqs.~(\ref{EquationAB}) yields
for each values of $J\in \{0,1,2,\ldots \}$ and $|M|\leq J$, the following decoupled system for the coefficients $(\mathcal{A}_{JM},\mathcal{B}_{JM}) := \left.\{ (A_{JM}{}^L,B_{JM}{}^L) \} \right|_{L=|J-\ell|,\ldots,J+\ell}$:
\begin{subequations}\label{EqGP}
\begin{align}
    i\lambda A_{JM}{}^L &=\left(\hat{\mathcal{H}}_{L}^{(0)}-E\right)B_{JM}{}^L, \label{EqGP1}\\
    i\lambda B_{JM}{}^L &=\left(\hat{\mathcal{H}}_{L}^{(0)}-E\right)A_{JM}{}^L + 2Q_{JM}{}^L,\label{EqGP2}
\end{align}
\end{subequations}
where $\mathcal{H}_{L}^{(0)}$ is defined similarly as in the previous subsection, that is
\begin{align}
    \mathcal{\hat H}_L^{(0)} := -\Delta_L + \Delta^{-1}_{0}\left( |\sigma_{\ell}^{(0)}|^2 \right),
\label{Eq:HL}
\end{align}
where $\Delta_L$ is defined as in Eq.~(\ref{Eq:DeltaJ}) (with $J$ replaced with $L$). Furthermore, the system decouples into two subsystems: the even-parity sector which contains $L = |J-\ell|,|J-\ell|+2,\ldots J+\ell$ and has non-trivial coefficients $Q_{JM}{}^L$ given in Eq.~(\ref{Eq:QJML}) and the odd-parity sector with $L = |J-\ell|+1,|J-\ell|+3,\ldots, J+\ell-1$ which has vanishing $Q_{JM}{}^L$.

For $\ell=1$ one has
\begin{equation}
C^{J0}_{J-1,0,1,0} = \sqrt{\frac{J}{2J-1}},\qquad
C^{J0}_{J+1,0,1,0} = -\sqrt{\frac{J+1}{2J+3}},
\end{equation}
and the system~(\ref{EqGP}) reduces to the system~(\ref{Eqell1}, \ref{Eq:LinSyst2}) in the previous subsection. Explicit examples of the resulting perturbation equations for $\ell=0,1,2$ are shown in Appendix~\ref{App:ExplicitLinSys}. In Appendix~\ref{App:Decoupling} we show that the perturbed evolution equation~(\ref{Eq:PerturbationEq}) similarly decouples into the different $JM$ and parity modes. Furthermore, we prove in that appendix that only purely oscillatory modes with purely imaginary $\lambda$ can occur in the odd-parity sector.

%%%%%%%%%%%%%%%%%%%%%%%%%%%%%%%%%%%%%%%
\section{Properties of the solutions of the linearized system}
\label{Sec:Properties}
%%%%%%%%%%%%%%%%%%%%%%%%%%%%%%%%%%%%%%%

Before numerically solving the linearized system~(\ref{EqGP}), in this section we discuss some important general properties of its solutions. For the following, we assume that $\chi_0$ is real-valued.

\subsection{Quadruple symmetry}
\label{Sec:IVA}
 
When $\chi_0$ is real, it is simple to see that a solution $(\lambda,\mathcal{A},\mathcal{B})$ of the system~(\ref{EquationAB}) gives rise to the three other solutions $(\bar{\lambda},\bar{\mathcal{A}},-\bar{\mathcal{B}})$, $(-\lambda,\mathcal{A},-\mathcal{B})$, $(-\bar{\lambda},\bar{\mathcal{A}},\bar{\mathcal{B}})$. Likewise, any solution $(\lambda, A_{JM}{}^L, B_{JM}{}^L)$ of the system~(\ref{EqGP}) yields the other three solutions $(-\lambda, A_{JM}{}^L, -B_{JM}{}^L)$, $(\bar\lambda, \bar{A}_{JM}{}^L, -\bar{B}_{JM}{}^L)$, and $(-\bar{\lambda}, \bar{A}_{JM}{}^L, \bar{B}_{JM}{}^L)$. This means that the eigenvalues come in pairs $(\lambda, -\lambda)$ if they are real or purely imaginary, and in quadruples $(\lambda, -\lambda, \bar\lambda, -\bar\lambda)$ otherwise.

\subsection{Stationary modes}
\label{Sec:IVB}

Next, we analyze the presence of stationary modes, that is, solutions of the system~(\ref{EqGP}) with $\lambda = 0$. In this case, Eq.~(\ref{EqGP1}) implies that $B_{JM}{}^L$ must be an eigenfunction of $\hat{\mathcal{H}}_{L}^{(0)}$ with eigenvalue $E$. When $L = \ell$, we know that $B_{JM}{}^\ell = \sigma_{\ell}^{(0)}$ satisfies this condition, because of the background equations~(\ref{StationaryProblem}). {\em A priori} it seems possible that $E$ also lies in the point spectrum of $\hat{\mathcal{H}}_{L}^{(0)}$ for values of $L$ different from $\ell$; however we do not pursue this issue further in this article. When $\lambda=0$, Eq.~(\ref{EqGP2}) leads to a homogeneous equation for $A_{JM}{}^L$. In this article, we only consider the trivial solution $A_{JM}{}^L = 0$, leaving open the problem of the existence of nontrivial solutions.

Summarizing, for given values of $\ell$, $J\in \{ 0,1,\ldots, 2\ell \}$ and $|M|\leq J$, there is a one-parameter family of zero modes of the form\footnote{Note that in view of the orthogonality property of the tensor spherical harmonics, the orthogonality conditions~(\ref{Eq:OrthoBis}) are satisfied.}
\begin{equation}
\label{Eq:Rel_Eing_Sta}
(A_{JM}{}^L, B_{JM}{}^L) = \Gamma_{JM}(0, S_{JM}{}^L),
\end{equation}
with $\Gamma_{JM}$ an arbitrary complex constant and where the fields $S_{JM}{}^L$ are zero except when $L=\ell$ in which case it is equal to $\sigma_{\ell}^{(0)}$. This leads to a multivalue family of stationary solutions of the linearized equations~(\ref{Eq:PerturbationEq}) which is of the form
\begin{equation}
\chi(t,\vec{x}) = \sigma_{\ell}^{(0)}(r)\sum\limits_{J=0}^{2\ell} \sum\limits_{M=-J}^J \left[ \Gamma_{JM} Y^{JM}{}_{\ell\ell}(\vartheta,\varphi) - c.c. \right],
\end{equation}
where $c.c.$ denotes complex conjugation. When $\ell=0$ there is only one mode which describes a change in amplitude of the background field, as discussed in~\cite{Roque:2023sjl}. However, when $\ell > 0$, there are $(2\ell+1)^2$ of these modes and, except the one with $J=0$, all these modes have an angular dependency which is different from the one of the background solution. As an example, consider $\ell$-boson stars with $\ell=1$. Then, we have stationary modes with angular dependency
\begin{eqnarray}
Y^{10}{}_{11} &=& \frac{1}{\sqrt{2}} \left[ Y^{1-1}\xi^{11} + Y^{11}\xi^{1-1} \right] 
\nonumber\\
 &=& 
-\frac{3}{8\pi}\sin\vartheta\left[ \cos\varphi\hat{e}_x + \sin\varphi\hat{e}_y \right].
\end{eqnarray}
Note that the zero modes discussed here belong to the even-parity sector when $J$ is even and to the odd-parity sector otherwise.

We conjecture that these modes lead to nonspherical stationary deformations of the $\ell$-boson stars.

\subsection{General properties and connection with the second variation of the energy functional}
\label{Sec:IVC}

Multiplying both sides of Eq.~(\ref{EquationA}) from the left with $\mathcal{B}^*$ and integrating yields
\begin{equation}
i\lambda(\mathcal{B},\mathcal{A}) = (\mathcal{B},(\hat{\mathcal{H}}_0 - E)\mathcal{B}),
\label{Eq:ProdBA}
\end{equation}
where $(\cdot,\cdot)$ refers to the $L^2$-scalar product defined in~(\ref{Eq:L2Product}). Likewise, multiplying both sides of Eq.~(\ref{EquationB}) from the left with $\mathcal{A}^*$ and integrating gives
\begin{eqnarray}
i\lambda(\mathcal{A},\mathcal{B}) &=& (\mathcal{A},(\hat{\mathcal{H}}_0 - E)\mathcal{A})
 + 2(\chi_0^T\mathcal{A},\Delta^{-1}[\chi_0^T\mathcal{A}])
\nonumber\\
 &=& \delta^2\mathcal{E}[\mathcal{A}_R] + \delta^2\mathcal{E}[\mathcal{A}_I],
\label{Eq:ProdAB}
\end{eqnarray}
where $\mathcal{A}_R$ and $\mathcal{A}_I$ refer to the real and imaginary parts of $\mathcal{A}$, respectively and $\delta^2\mathcal{E}[\mathcal{A}_R]$ denotes to the second variation~(\ref{Eq:SecondVariationchi0}) evaluated at $\delta u = \mathcal{A}_R$ with fixed particle numbers $N_j$.

Similar to the analysis in our previous work~\cite{Roque:2023sjl}, several interesting features can be inferred from Eqs.~(\ref{Eq:ProdBA}, \ref{Eq:ProdAB}). For this, we first note that the right-hand sides of these equations are real, which implies that
\begin{equation}
-\lambda^2\left| (\mathcal{A},\mathcal{B}) \right|^2 \in \Real.
\end{equation}
Hence, either $\lambda^2$ is real or $\mathcal{A}$ is orthogonal to $\mathcal{B}$. Taking into account the quadruple symmetry, we may consider the following cases:
\begin{enumerate}
\item[(i)] $\lambda=0$: These are the zero modes discussed previously.
\item[(ii)] $\lambda_R > 0$ and $\lambda_I = 0$: In this case we can assume that $\mathcal{A} = \mathcal{A}_R$ is real and $\mathcal{B} = i\mathcal{B}_I$ is purely imaginary. Eliminating $i\lambda\mathcal{A}$ on the left-hand side of Eq.~(\ref{Eq:ProdAB}) using Eq.~(\ref{EquationA}), one finds
\begin{equation}
-(\mathcal{B}_I,(\hat{\mathcal{H}}_0 - E)\mathcal{B}_I) = \delta^2\mathcal{E}[\mathcal{A}_R].
\label{Eq:IdentityRealLambda}
\end{equation}
Below, we will use this identity to eliminate the possibility of having unstable modes with arbitrary high values of $J$.
\item[(iii)] $\lambda_R = 0$ and $\lambda_I > 0$: In this case one can choose both $\mathcal{A}$ and $\mathcal{B}$ to be real, and one obtains instead of Eq.~(\ref{Eq:IdentityRealLambda}),
\begin{equation}
(\mathcal{B}_R,(\hat{\mathcal{H}}_0 - E)\mathcal{B}_R) = \delta^2\mathcal{E}[\mathcal{A}_R].
\label{Eq:IdentityComplexLambda}
\end{equation}
\item[(iv)] $\lambda_R > 0$ and $\lambda_I > 0$: In this case $(\mathcal{A},\mathcal{B}) = 0$ and it follows from Eq.~(\ref{Eq:ProdAB}) that $\chi_0$ is a saddle point of $\mathcal{E}$, provided that $\mathcal{E}[\mathcal{A}_R]\neq 0$.
\end{enumerate}
In terms of the decomposition~(\ref{Eq:Descomp}) into tensor spherical harmonics, the scalar product $(\mathcal{A},\mathcal{B})$ reads
\begin{equation}
(\mathcal{A},\mathcal{B}) 
 = \sum\limits_{JM} \left[ (\mathcal{A}_{JM},\mathcal{B}_{JM})_{\text{even}}
 + (\mathcal{A}_{JM},\mathcal{B}_{JM})_{\text{odd}}
 \right],
\end{equation}
with
\begin{align}
& (\mathcal{A}_{JM},\mathcal{B}_{JM})_{\text{even, odd}} &
\nonumber\\
 := &\sum_{\substack{L=|J-\ell| \\ \text{ $J+\ell-L$ even,odd}}}^{J+\ell} \int\limits_0^\infty \overline{A_{JM}{}^L(r)} B_{JM}{}^L(r) r^2 dr
\end{align}
denoting the corresponding products for the $JM$ modes in the even and odd parity sectors. A similar decomposition can be performed for the previous equations in this subsection; for instance Eq.~(\ref{Eq:ProdAB}) yields
\begin{equation}
i\lambda(\mathcal{A}_{JM},\mathcal{B}_{JM})_{\text{even}} = \delta^2\mathcal{E}_{JM,even}[\mathcal{A}_{JM}],
\label{Eq:ProdABEven}
\end{equation}
where $\delta^2\mathcal{E}_{JM,even}[\mathcal{A}_{JM}]$ is computed in Appendix~\ref{App:ModeDecompEnergyFunc}. In the next section, we shall use Eq.~(\ref{Eq:ProdABEven}) to check numerically that $i\lambda(\mathcal{A}_{JM},\mathcal{B}_{JM})_{\text{even}}$ is real.

%%%%%%%%
\subsection{Real eigenvalues}
\label{Sec:IVD}
%%%%%%%%%

In contrast to spherically symmetric perturbations ($J=0$) discussed in our previous work~\cite{Roque:2023sjl}, in the next section we will see that nonzero real eigenvalues are possible when $J > 0$. Recall that in this case, the ansatz~(\ref{Eq:SigmaAnsatz}) reduces to $\chi(t,\vec{x}) = e^{\lambda t}\left[ \mathcal{A}_R(\vec{x}) + i\mathcal{B}_I(\vec{x})\right]$, such that one needs to make sure that $\mathcal{A}_R$ and $\mathcal{B}_I$ are not both zero for the corresponding mode to be physically relevant. However, due to the linearity of the system~(\ref{EquationAB}), it is clear that this can always be achieved by multiplying $\mathcal{A}$ and $\mathcal{B}$ with a phase factor if necessary, such that it is sufficient to check the standard eigenvector condition that $\mathcal{A}$ and $\mathcal{B}$ are not both zero.

\subsection{Non-existence of unstable modes for sufficiently large values of $J$}
\label{Sec:IVE}

Finally, in this section we prove that for modes with large enough values of the total angular momentum $J$, the second variation of the energy functional given by equation (\ref{Eq:SecondVariationchi0}) is positive definite. As we show below, this implies through Eq.~(\ref{Eq:IdentityRealLambda}) that there cannot exist unstable modes with large $J$. This reduces the stability problem to the analysis of a finite number of $J$.

The proof is based on the following estimate which is proven in Appendix~\ref{App:EstimateSecondVar}:
\begin{eqnarray}
\delta^2\mathcal{E} &\geq& \frac{1}{2}(\nabla\delta u,\nabla \delta u) + (\delta u, [U_0 - E]\delta u) 
\nonumber\\
 &-& C_1 \|\delta u/f\|_2^2,
\label{Eq:EstimateSecVar}
\end{eqnarray}
where $f$ is a positive function of $r$ which will be determined shortly, $C_1 > 0$ a positive constant depending on $f$ and $U_0 := \triangle^{-1}(|\chi_0|^2)$ is the gravitational potential of the background configuration. To show that $\delta^2\mathcal{E}$ is positive definite for large enough $J$ we expand $\delta u$ in terms of the tensor spherical harmonics:
\begin{equation}
\delta u = \sum_{JLM} h_{JM}{}^L(r) Y^{JM}_{L\ell} (\vartheta, \varphi),
\end{equation}
with coefficients $h_{JM}{}^L$ depending on $r$. Substituting this in the right-hand side of Eq.~(\ref{Eq:EstimateSecVar}) and discarding the quadratic terms in the derivatives of $h_{JM}{}^L$ yields
\begin{align}
   \delta^2\mathcal{E} &\geq \sum_{JLM}\left\{\int_{0}^{\infty} |h_{JM}{}^L(r)|^2\left[ \frac{L(L + 1)}{2} - \frac{C_1 r^2}{f(r)^2}\right] dr\right.
\nonumber\\
   &+ \left. \int_{0}^{\infty} |h_{JM}{}^L(r)|^2 [U_0(r) - E]r^2 dr \right\}.
\label{Eq:LargeJEstimate}
\end{align}
Consider first the integral on the second line, whose integrand contains the function $g(r) := [U_0(r) - E]r^2$. Since $U_0$ is regular at the center, one has $g(0) = 0$, whereas $g(r)$ is positive for large enough $r$ since $E$ is negative. Together with the fact that $U_0$ is continuous, this implies that $g(r)\geq C_2$ for all $r\geq 0$, for some (negative) constant $C_2$. Next, choose $f(r) := \sqrt{1 + r^2}$ which implies that $r^2/f(r)^2 \leq 1$ for all $r\geq 0$. Using these properties, the estimate~(\ref{Eq:LargeJEstimate}) yields
\begin{equation}
\delta^2 \mathcal{E} \geq \sum_{JLM} \int_{0}^{\infty} |h_{JM}{}^L(r)|^2\left[ \frac{L(L + 1)}{2} - C_1 + C_2\right] dr.
\end{equation}
Therefore, $\delta^2\mathcal{E}$ is positive definite if $h_{JM}{}^L$ vanishes identically for all $L$ with $L(L + 1)/2 - C_1 + C_2\leq 0$. In particular, it follows that $\delta^2\mathcal{E}_{JM,even}$ is positive definite for $J$ large enough, such that $L:=|J-\ell|$ satisfies $L(L+1) > 2(C_1 - C_2)$. 

Finally, we prove that this property implies the absence of unstable modes for large enough values of $J$. We do this by contradiction. Consider first case (iv) for which $\lambda_R,\lambda_I > 0$ and $(\mathcal{A},\mathcal{B}) = 0$. In this case, Eq.~(\ref{Eq:ProdAB}) and the positivity of $\delta^2\mathcal{E}$ would imply that $\mathcal{A}_R = \mathcal{A}_I = 0$ which also implies that $\mathcal{B} = 0$ according to Eqs.~(\ref{EquationAB}). The other case in which an instability could appear is case (ii).  Here, a contradicion arises by observing that the right-hand side of Eq.~(\ref{Eq:IdentityRealLambda}) is positive definite, whereas the left-hand side is negative definite for large enough values of $J$, as can be shown using arguments similar to the ones following Eq.~(\ref{Eq:LargeJEstimate}).

%%%%%%%%%%%%%%%%%%%%%%%%%%%%%%%%%%%%%%%
\section{Numerical results}
\label{Sec:Numerical}
%%%%%%%%%%%%%%%%%%%%%%%%%%%%%%%%%%%%%%%
%%%%%%%%%
{\captionsetup[table]{name=FIG. 1/TABLE}
\begin{table*}[tb]
\raggedright
%\centering
\sbox\tmpbox{%
    \renewcommand{\tabcolsep}{10pt}
    \renewcommand {\arraystretch}{1.5}
    \centering
    \begin{tabular}[b]{ ||c@{\hskip .1 in} |c @{\hskip .2 in} l@{\hskip .1 in}|| c@{\hskip .1 in} | c @{\hskip .2 in} l@{\hskip .1 in}||}
      \toprule
       & \multicolumn{2}{|c||}{$\lambda \; [N^2/t_c]$} & & \multicolumn{2}{|c||}{$\lambda \; [N^2/t_c]$}\\[0.05cm]
       & \multicolumn{2}{|c||}{Components} & & \multicolumn{2}{|c||}{Components}\\[-0.1cm]
       $J$ & Real & Imaginary & $J$ & Real & Imaginary \\
      \midrule
      \multirow{3}{*}{$0$} & $0$ & $1.05201\times10^{-5}$ & \multirow{3}{*}{$4$} & $0$ & $0.11368481$\\[-0.16cm]
        & $0$ & $0.06385090$ &                      & $0$ & $0.16030883$ \\[-0.16cm]
        & $0$ & $0.14328038$ &                      & $0$ & $0.18097544$ \\[0.1cm]
        %& $0$ & $0.17565512$ &                      & $0$ & $0.19757833$ \\[0.1cm]
      %
      \multirow{3}{*}{$1$} & $0$ & $0.00086601$ & \multirow{3}{*}{$5$} & $0$ & $0.15468802$ \\[-0.16cm]
                           & $0$ & $0.05499307$ &                      & $0$ & $0.18203923$  \\[-0.16cm]
                           & $0$ & $0.08194250$ &                      & $0$ & $0.19756177$  \\[0.1cm]
     \multirow{3}{*}{$2$} & $0$ & $2.11418\times10^{-6}$  & \multirow{3}{*}{$6$}  & $0$ & $0.18105790$ \\[-0.16cm]
     & $0$ & $0.09613376$  &  & $0$ & $0.19765391$ \\[-0.16cm]
     & $0$ & $0.10590972$  &  & $0$ & $0.20832581$  \\[0.1cm]
    \multirow{3}{*}{$3$}  & $0$ & $0.05557199$ & \multirow{3}{*}{$7$}  & $0$ & $0.19756669$ \\[-0.16cm]
                          & $0$ & $0.13196275$ &                       & $0$ & $0.20833027$  \\[-0.16cm]
                          & $0$ & $0.15373565$ &                       & $0$ & $0.21570444$  \\[0.1cm]
      \bottomrule
    \end{tabular}
  }
\renewcommand*{\arraystretch}{0}
\begin{tabular*}{0.8\textwidth}{@{\extracolsep\fill} p{70mm}@{} p{70mm}@{}}
\includegraphics[width=\linewidth]{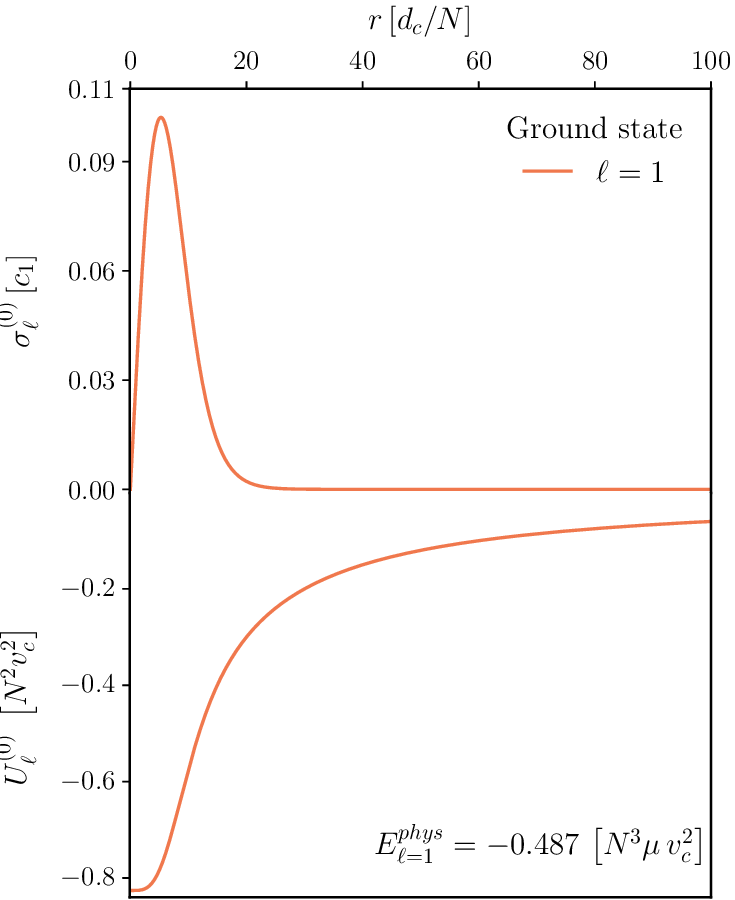}
& \usebox\tmpbox \\
\end{tabular*}
\caption{The left panel shows the background profiles' wave function $\sigma_{\ell}^{(0)}$ and gravitational potential $U_{\ell}^{(0)}$ for the $(\ell=1)-$boson star whose bosons lie in the ground state $n=0$ and have an energy $E^{phys}_{\ell=1}=-0.487 \,\left[N^3 \mu\, v_c^2\right]$. The right panel shows a table with the first three eigenvalue pairs $(\lambda, -\lambda)$ corresponding to perturbations with total angular momentum numbers $J=1,2,\dots,7$ of the configuration shown in the left panel. We found only purely oscillatory modes. Here $c_{1}$ refers to the constant $N^2/\sqrt{4\pi d_c^{3}}$ appearing in Eq.~(\ref{AdVar}), where $t_c$ and $v_c$ are also defined.}\label{Fig/Tab1}
\end{table*}
}

%%%%%
In section~\ref{Sec:LinSys} we derived the mode equations for the nonrelativistic $\ell$-boson stars. We first considered radial perturbations and then extended the methodology to the non-radial case for $\ell=1$ (system~(\ref{Eq:LinSyst}) or equivalently~(\ref{Eqell1})). In subsection~\ref{Sec:LinSyslbosonstars} we generalized the method to arbitrary values of $\ell$ (see system~(\ref{EqGP}) and Appendix~\ref{App:ExplicitLinSys} for some examples). In Appendix~\ref{App:Decoupling} and the previous section, some general properties of the linearized system were established. In particular, it was proven that unstable modes cannot arise in the odd-parity sector nor in the even-parity sector with high values of $J$, thus reducing the problem to a finite number of decoupled systems. For this reason, we will focus on the even-parity sector for what follows.

We start in the next subsection with a short description of our numerical implementation and subsequently, we discuss our main results regarding the eigenvalues of~(\ref{EqGP}).

%In particular, it was shown that this system can be decoupled into two sectors: an even-parity sector with non-zero coefficients $Q_{JM}{}^L$ and an odd-parity sector with zero coefficients $Q_{JM}{}^L$ (see ). 
%We prove in appendix~\ref{App:Decoupling} that these decouples come from a decoupling in the perturbated evolution equation and that the odd-parity sector has a unitary evolution, contributing only with oscillatory modes, i.e., purely imaginary $\lambda$ values. This means that any unstable modes of system~(\ref{EqGPN}), i.e., modes with $\lambda$ values with a non-zero real component, must come from the even-parity sector. 
%Therefore, from now on, we will only focus on this sector.
%Here we present our numerical implementation and main results corresponding to the numerical analysis of the even-parity sector.

%%%%%
\subsection{Implementation}
%%%%

%
%%%%%%%%%%%%%%%%%%
{\captionsetup[table]{name=FIG. 2/TABLE}
\begin{table*}[th]
\raggedright
\sbox\tmpbox{%
    \renewcommand{\tabcolsep}{9pt}
    \renewcommand {\arraystretch}{1.53}
    \centering
    \begin{tabular}[b]{||@{\hskip .015 in}l@{\hskip .015 in}| @{\hskip .03 in}l@{\hskip .015 in} c@{\hskip .015 in}| @{\hskip .05 in}l@{\hskip .05 in} c@{\hskip .015 in}| @{\hskip .05 in}l@{\hskip .05 in} c@{\hskip .015 in}||}
      \toprule
       & \multicolumn{2}{c}{$\ell=0, \, \lambda \; [N^2/t_c]$} & \multicolumn{2}{l}{$\ell=2, \, \lambda \; [N^2/t_c]$}& \multicolumn{2}{l||}{$\ell=3, \, \lambda \;[N^2/t_c]$} \\[-0.1cm]
       & \multicolumn{2}{l}{Components} & \multicolumn{2}{l}{Components} & \multicolumn{2}{l||}{Components}\\[-0.1cm]
       $J$ & Real & Imaginary & Real & Imaginary  & Real & Imaginary \\
      \midrule
      \multirow{3}{*}{$0$} & $0$ & $4.321\times10^{-6}$ & $0$ & $8.766\times10^{-6}$ & $0$ & $1.637\times10^{-5}$\\[-0.16cm]
                           & $0$ & $0.03412558$ & $0$ & $0.06408695$ & $0$ & $0.05903587$ \\[-0.16cm]
                           & $0$ & $0.06030198$ & $0$ & $0.16575905$ & $0$ & $0.16827468$ \\[0.1cm]
\multirow{3}{*}{$1$} & $0$ & $0.00049999$ & $0$ & $0.00111802$ & $0$ & $0.00132286$\\[-0.16cm]
                           & $0$ & $0.05555078$ & $0$ & $0.05842642$ & $0$ & $0.05925506$ \\[-0.16cm]
                           & $0$ & $0.06676208$ & $0$ & $0.07777395$ & $0$ & $0.06609698$\\[0.1cm]
\multirow{3}{*}{$2$} & $0$ & $0.05285868$  & $4.32\times10^{-7}$ & $2.63\times10^{-15}$ & $7.99\times10^{-7}$ & $1.37 \times 10^{-14}$ \\[-0.16cm]
& $0$ & $0.06564270$  & $0.00644373$ & $-5.11\times10^{-13}$ & $0.00907756$ & $2.1\times10^{-12}$ \\[-0.16cm]
& $0$ & $0.07136165$  & $0$ & $0.10237935$ & $0$ & $0.09954769$  \\[0.1cm]
\multirow{3}{*}{$3$}  & $0$ & $0.06571493$ & $0$ & $0.05550210$ & $0.00363560$ & $0.05313153$ \\[-0.16cm]
& $0$ & $0.07135829$ & $0$ & $0.06147704$ & $0$ & $0.05910000$ \\[-0.16cm]
& $0$ & $0.07442765$ & $0$ & $0.07000546$ & $0.00085739$ & $0.05953874$ \\[0.1cm]
      \bottomrule
    \end{tabular}%
  }%
%%%%%%
\renewcommand*{\arraystretch}{0}
\begin{tabular*}{0.785\textwidth}{@{\extracolsep\fill} p{70mm}@{} p{70mm}@{}}
\includegraphics[width=\linewidth]{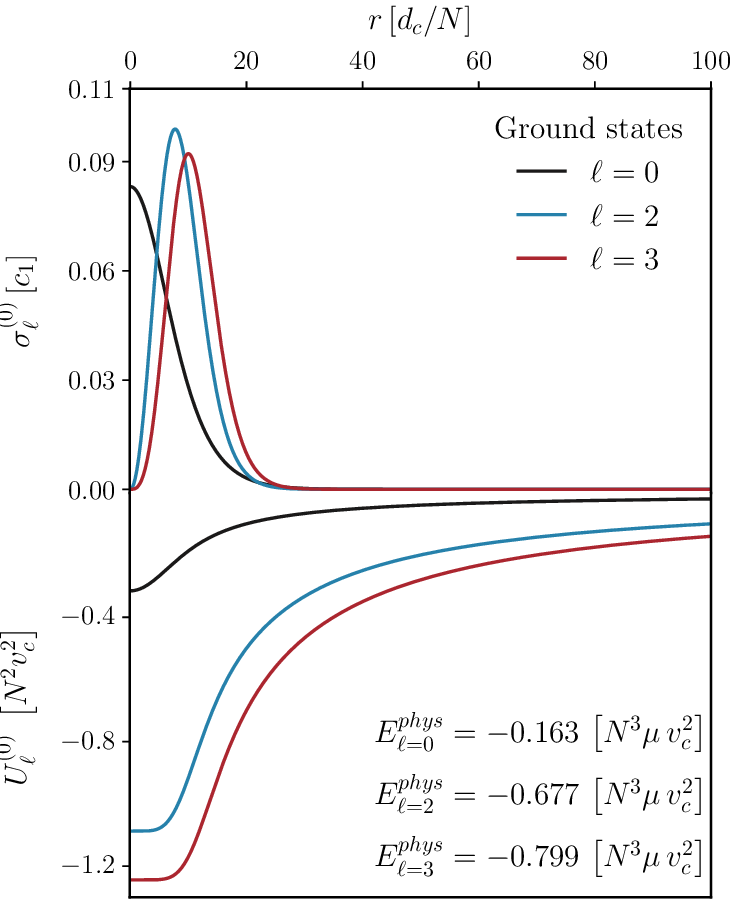}
& \usebox\tmpbox \\
\end{tabular*}
\caption{The left panel shows the background profiles' wave function $\sigma_{\ell}^{(0)}$ and gravitational potential $U_{\ell}^{(0)}$ for the $\ell=0, 2,3$ boson stars whose bosons lie in the ground state $n=0$ with an energy $E^{phys}_{\ell=1}=-0.163, -0.677, -0.799 \,\left[N^3\mu\, v_c^2\right]$, respectively. The right panel shows a table with the first three lowest eigenvalues of the perturbations with $J=0, 1, \dots, 3$ of the configuration shown in the left panel. We found only purely oscillatory modes for $\ell=0$, while the remaining configurations have unstable modes, as can be appreciated from the table. Here $c_{1}$ refers to the constant $N^2/\sqrt{4\pi d_c^{3}}$ appearing in Eq.~(\ref{AdVar}).}
\label{Fig/Tab2}
\end{table*}
}
%%%%%%%%%
%

Our methodology is similar to the one implemented in our previous paper~\cite{Roque:2023sjl}. The background profiles are computed by solving the non-linear eigenvalue problem~(\ref{StationaryProblem}) with the ansatz~(\ref{Eq:chi0lboson}). Introducing the shifted potential $u^{(0)}(r):= E - \triangle^{-1}_{0}\left(\abs{\sigma_{\ell}^{(0)}}^2\right)$, the equation~(\ref{StationaryProblem}) is reduced to the system (41) in Ref.~\cite{Roque:2023sjl}. Since the main goal of this article consists in the study of the linearized system~(\ref{EqGP}), we refer the reader to Ref.~\cite{Roque:2023sjl} for a detailed analysis of the construction of the background configurations. In the following, we assume that we have already computed the numerical background profiles $\sigma_\ell^{(0)}(r), u^{(0)}(r)$.

Introducing the change of variables ${A_{JM}}^L={a_{JM}}^L/r$, ${B_{JM}}^L={b_{JM}}^L/r$ in~(\ref{EqGP}) one obtains
\begin{subequations}\label{EqGPN}
\begin{align}
    b{''}_{JM}{}^L-U_{\text{eff}}{}^L b_{JM}{}^L&=-i\lambda a_{JM}{}^L,\\
    a{''}_{JM}{}^{L}-U_{\text{eff}}{}^L a_{JM}{}^L-2q_{JM}{}^{L}&=
    -i\lambda b_{JM}{}^L,
\end{align}
\end{subequations}
where a prime denotes differentiation with respect to $r$, $U_{\text{eff}}{}^{L}(r):=L(L+1)/r^2-u^{(0)}(r)$ is an effective potential, and the function $q_{JM}{}^L$ is defined by
\begin{align}
&q_{JM}{}^L(r):= \sigma_\ell^{(0)}(r)\sum\limits_{L'=|J-\ell|}^{J+\ell}
\frac{\sqrt{(2L+1)(2L'+1)}}{2J+1} C^{J0}{}_{L0\ell 0} 
\nonumber\\
 &\times  C^{J0}{}_{L'0\ell 0} 
 \bigg(\frac{d^2}{dr^2}-\frac{J(J+1)}{r^2}\bigg)^{-1}\left[\sigma_\ell^{(0)} a_{JM}{}^{L'}\right](r), \label{Eq:QJMLN}
\end{align}
with the operator $\bigg(\frac{d^2}{dr^2}-\frac{J(J+1)}{r^2}\bigg)^{-1}=r\triangle^{-1}_{J}(r^{-1})$ denoting the inverse of the operator $r\triangle_{J}(r^{-1})$ with homogeneous Dirichlet conditions at $r = 0$ and $r = \infty$. Note that the system~(\ref{EqGPN}), like the system~(\ref{EqGP}), is independent of the total magnetic quantum number $M$, and hence we do not need to specify it.

To solve the system~(\ref{EqGPN}) we need two boundary conditions for each equation. To determine these, one can study (heuristically) the dominant terms of the perturbed system near the origin and infinity. Using the fact that $J = 0, 1, 2, \ldots$ and $L = |J-\ell|,|J-\ell|+2,\ldots, J+\ell$, and that the background solution behaves as $\sigma_\ell^{(0)}(r)\sim r^{\ell}$, one finds that the dominant terms at the center stem from the centrifugal terms $L(L+1)/r^2$ in the effective potential. Consequently, the regular solution at the center behaves as $(a_{JM}{}^L, b_{JM}{}^L)\sim (r^{L+1}, r^{L+1})$ (see Appendix~\ref{App:FirstOrderFormulation} for further details). This leads to the following boundary conditions for all $\ell\geq 0$ at the origin:
\begin{subequations}
\begin{align}
a_{JM}{}^L(r=0)=0,\qquad b_{JM}{}^L(r=0)=0.
\label{PertBC}
\end{align}
In the asymptotic region $\sigma_\ell^{(0)}$ decays exponentially and $u^{(0)}(r)\to E$. Demanding that the fields $(a_{JM}{}^L, b_{JM}{}^L)$ decay at infinity, one requires that
\begin{align}
\lim\limits_{r\to\infty} a_{JM}{}^L(r)=0,\qquad \lim\limits_{r\to\infty} b_{JM}{}^L(r)=0.
\end{align}
\end{subequations}

In order to solve numerically the system~(\ref{EqGPN}) using the previous Dirichlet boundary conditions we proceed as follows. First, we computed the background profiles $\sigma_\ell^{(0)}, u^{(0)}$ and represent these, as well as the perturbed fields $a_{JM}{}^{L}, b_{JM}{}^{L}$, in terms of Chebyshev polynomials. The different operators e.g., derivative and its inverse are discretized using a standard spectral method (see, e.g., Ref.~\cite{trefethen2000spectral}), which leads to a finite-dimensional eigenvalue problem. For details of the numerical discretization procedure, we refer the reader to subsection IVA in our previous paper Ref.~\cite{Roque:2023sjl}. 

%As was mentioned earlier, the system~(\ref{EqGP}) can be decoupled into two subsystems: the even-parity sector with non-trivial coefficients $Q_{JM}{}^L$ and the odd-parity sector which has vanishing $Q_{JM}{}^L$ (see appendix~\ref{App:ExplicitLinSys} for some examples). In appendix~\ref{App:Decoupling} we prove that these decouples come from a decoupling in the evolution perturbation's equation and that the odd-parity sector has a unitary evolution, contributing only with oscillatory modes, i.e., purely imaginary $\lambda$ values. This means that, if the system~(\ref{EqGPN}) has unstable modes, i.e., $\lambda$ values with a non-null real component, they would come from their even-parity sector. The reason why from now on, we only focus on this sector.

The discrete version of the system~(\ref{EqGPN}) can be written as:
\begin{widetext}
\begin{gather}\label{Syst}
 \begin{pmatrix} \mathsf{0} & \mathbb{D}^{2}-\mathbb{U}_{\ell J} \\ \mathbb{D}^{2}-\mathbb{U}_{\ell J}-2\Sigma_{\ell}\mathbb{Z}_{\ell J}(\mathbb{D}^{2}-\mathbb{V}_{J})^{-1}\Sigma_{\ell}& \mathsf{0}\end{pmatrix}\begin{pmatrix} \mathsf{a}_{JM}{} \\ \mathsf{b}_{JM}{}\end{pmatrix}
 =
 -i\lambda
   \begin{pmatrix} \mathsf{a}_{JM}{} \\ \mathsf{b}_{JM}{}\end{pmatrix},
\end{gather}
\end{widetext}
where here $\mathsf{0}$ represents the $c_J(\mathsf{N}-1)\times c_J(\mathsf{N}-1)$ zero matrix, with $\mathsf{N}$ the number of Chebyshev points distributed as $x_j=\cos{(j\pi/\mathsf{N})}, j=0,1,\dots,\mathsf{N}$. The constant $c_J$ is defined as $c_J := J+1$ for $J<\ell$ and as $c_J := \ell+1$ when $J\geq\ell$, and it corresponds to the number of possible values of $L$ with non-trivial coefficients $Q_{JM}{}^L$ for a given tuple $(\ell, J)$. $\mathbb{D}^{2}, \mathbb{U}_{\ell J}, \Sigma_{\ell}$ and $\mathbb{V}_{J}$ are $c_J(\mathsf{N}-1)\times c_J(\mathsf{N}-1)$ matrices whose diagonal contain $c_J$ blocks of smaller $(\mathsf{N}-1)\times (\mathsf{N}-1)$ matrices,
\begin{subequations}
\begin{align}
\mathbb{D}^{2}&=\textbf{diag}\left(\mathbb{\tilde{D}}_{\mathbb{N}}^{2}, \mathbb{\tilde{D}}_{\mathbb{N}}^{2}, \dots, \mathbb{\tilde{D}}_{\mathbb{N}}^{2} \right),\\
\mathbb{U}_{\ell J}&=\textbf{diag}\left(U_{\text{eff}}{}^{|J-\ell|}, U_{\text{eff}}{}^{|J-\ell|+2},\dots,  U_{\text{eff}}{}^{J+\ell}\right),\\
\Sigma_{\ell}&=\textbf{diag}\left(\Sigma_\ell^{(0)}, \Sigma_\ell^{(0)}, \dots, \Sigma_\ell^{(0)} \right),\\
\mathbb{V}_{J}&=\textbf{diag}\left(V_J, V_J, \dots, V_J \right).
\end{align}
\end{subequations}
The matrix block $\mathbb{\tilde{D}}_{\mathbb{N}}^{2}$ corresponds to the discrete representation of the second derivative operator with implemented Dirichlet conditions. For details of its construction we refer the reader to Refs.~\cite{trefethen2000spectral, Roque:2023sjl}.
The blocks $\Sigma_\ell^{(0)}, V_A$ and $U_{\text{eff}}{}^{L}$ are diagonal and are constructed as
\begin{subequations}
\begin{align}
\Sigma_\ell^{(0)}&=\textbf{diag}\left(\sigma_\ell^{(0)}(x_1), \sigma_\ell^{(0)}(x_2), \dots, \sigma_\ell^{(0)}(x_{\mathsf{N-1}})\right),\nonumber\\
V_A &= \textbf{diag}\left(\frac{A(A+1)}{x_1^{2}}, \frac{A(A+1)}{x_2^{2}}, \dots, \frac{A(A+1)}{x_{\mathsf{N-1}}^{2}}\right), \nonumber\\
U_{\text{eff}}{}^{L}&=V_L-\textbf{diag}\left(u^{(0)}(x_1), u^{(0)}(x_2), \dots, u^{(0)}(x_{\mathsf{N-1}})\right), \nonumber
\end{align}
\end{subequations}
where the subscript $A$ in $V_A$ can take the labels $J$ and $L$. The matrix $\mathbb{Z}_{\ell J}$ is non-diagonal, has dimension $c_J(\mathsf{N}-1)\times c_J(\mathsf{N}-1)$, and is obtained from
\begin{gather}
\mathbb{Z}_{\ell J}=\begin{pmatrix}
Z_{J L'=|J-\ell|}^{L=|J-\ell|} & Z_{J L'=|J-\ell|+2}^{L=|J-\ell|} &\cdots & Z_{J L'=J+\ell}^{L=|J-\ell|}\\[10pt]
Z_{J L'=|J-\ell|}^{L=|J-\ell|+2} & Z_{J L'=|J-\ell|+2}^{L=|J-\ell|+2} &\cdots & Z_{J L'=J+\ell}^{L=|J-\ell|+2}\\[5pt]
\vdots & \cdots & \ddots & \vdots\\[5pt]
Z_{J L'=|J-\ell|}^{L=J+\ell} & Z_{J L'=|J-\ell|+2}^{L=J+\ell} &\cdots & Z_{J L'=J+\ell}^{L=J+\ell}
 \end{pmatrix},
\end{gather}
where the blocks $Z_{J L'}^{L}$ are diagonals matrices of constant coefficients
\begin{align}
 Z_{J L'}^{L}&=\frac{\sqrt{(2L+1)(2L'+1)}}{2J+1} C^{J0}{}_{L0\ell 0} C^{J0}{}_{L'0\ell 0} \times \mathbb{I},
\end{align}
with $\mathbb{I}$ the identity matrix of dimension $(\mathsf{N}-1)\times (\mathsf{N}-1)$. The vector
\begin{align}
\begin{pmatrix} \mathsf{a}_{JM} \\ \mathsf{b}_{JM}\end{pmatrix}=\bigg(&a^{|J-\ell|}(x_1), \dots, a^{|J-\ell|}(x_{\mathsf{N}-1}), a^{|J-\ell|+2}(x_1),\nonumber\\ 
&\dots, a^{|J-\ell|+2}(x_{\mathsf{N}-1}), \dots\dots, a^{J+\ell}(x_{\mathsf{N}-1}),\nonumber\\
&b^{|J-\ell|}(x_1), \dots, b^{|J-\ell|}(x_{\mathsf{N}-1}), b^{|J-\ell|+2}(x_1),\nonumber\\ 
&\dots, b^{|J-\ell|+2}(x_{\mathsf{N}-1}), \dots\dots, b^{J+\ell}(x_{\mathsf{N}-1})\bigg)^{T},\nonumber
\end{align}
corresponds to the discrete representation of the eigenfields $r(A_{JM}{}^{L}, B_{JM}{}^{L})^T$. 

We solve the discrete eigenvalue problem~(\ref{Syst}) using the SciPy library~\cite{2020SciPy-NMeth} for $\mathsf{N}:=3r_{\star}/4$ Chebyshev points where $r_{\star}:=200(n+1)$ for the $n$'th excited state of the background solution represents the physical radius of the outer boundary of our numerical domain. Our code is publicly available in~\cite{Roque_On_the_radial_2023}. 

%%%%%
\subsection{Ground state in nonrelativistic $(\ell=1)$-boson stars}
%%%%%%

\setcounter{figure}{2}
\begin{figure*}[tb]
	\centering	
\includegraphics[width=\linewidth]{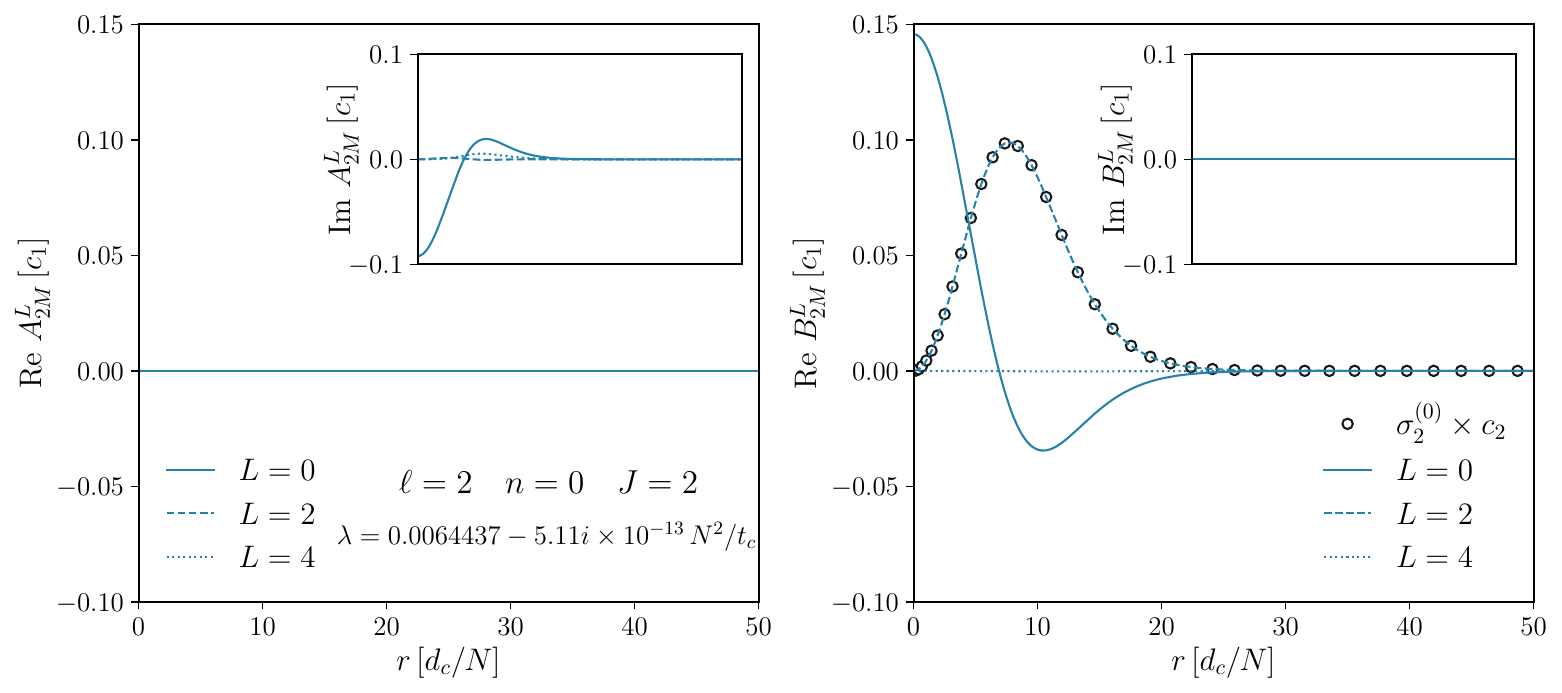}
	\caption{Components of the eigenfields $(\mathcal{A}_{J M}, \mathcal{B}_{J M})$ corresponding to the even-parity modes with $J=2$ of the configurations with $\ell=2$ and $n=0$. The profiles $A_{2M}^{L=0, 2, 4}$ (left panels) and $B_{2M}^{L=0, 2, 4}$  (right panels) are associated with real eigenvalues. In both cases, the main plots show the real parts, whereas the inset the imaginary ones. Note that the component $\mathcal{B}^{L=\ell}_{J M}$ is proportional to the background solution $\sigma_{2}^{(0)}$ (rescaled by a factor $c_2$ and shown with dark circles in the figure on the right panel), $\mathcal{B}_{JM}$ is real and $\mathcal{A}_{JM}$ is purely imaginary.}\label{Fig3}
\end{figure*}

\begin{figure*}[tb]
	\centering	
\includegraphics[width=\linewidth]{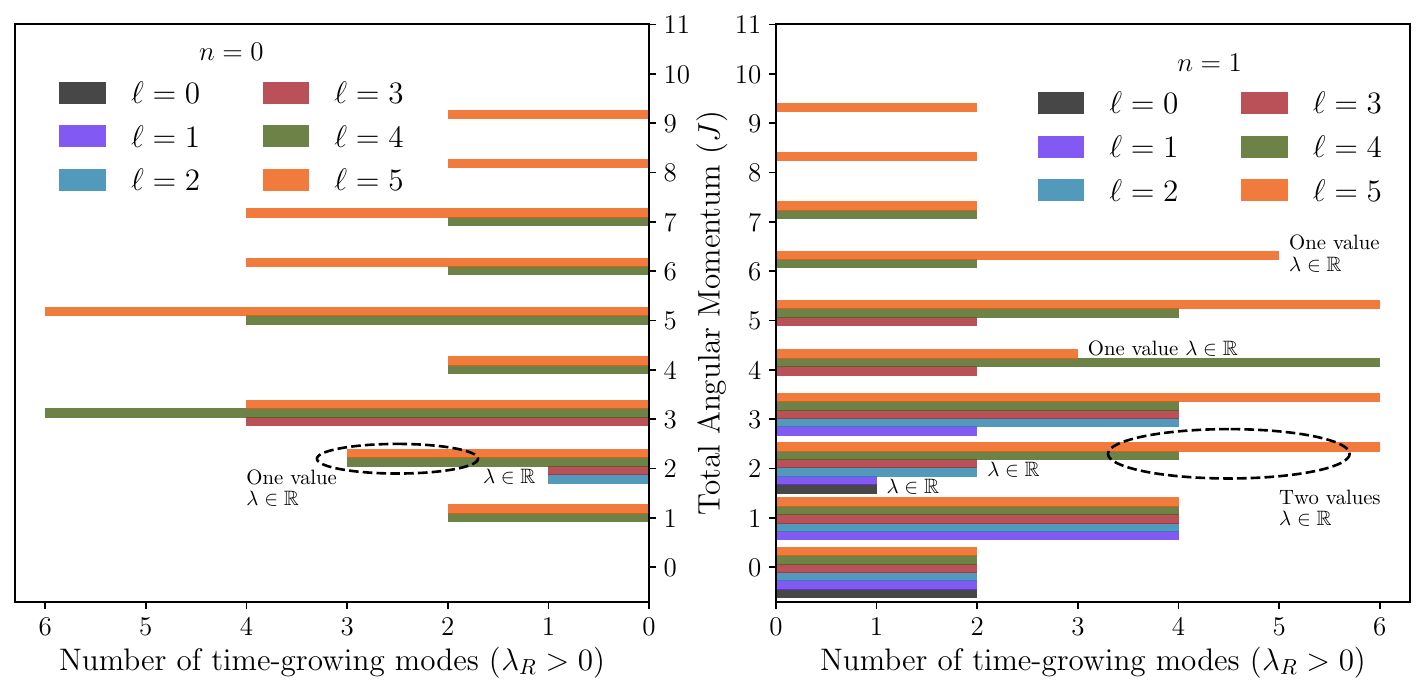}
	\caption{The relation between the total angular momentum $J$ and the number of time-growing modes i.e., eigenvalues with positive real parts $\lambda_{R}>0$. As a consequence of the symmetries discussed in subsection~\ref{Sec:IVA} we have one (two) growing mode (modes) for every real (complex) eigenvalue with $\lambda_{R}\neq 0$. The cases having real $\lambda$ are pointed out in the figures, e.g., for $n=0, \ell=2, 3$ and $J=2$  one has $\lambda\in \mathbb{R}$. Note that for large values of $J$ there are only oscillating modes. The left/right panel corresponds to configurations in the ground/first-excited states.}\label{Fig4}
\end{figure*}

We first proceed to study the linear stability of the ground state corresponding to a nonrelativistic $(\ell=1)$-boson star. This configuration is characterized by a radial scalar field profile $\sigma_{1}^{(0)}$ without nodes ($n=0$), whose gravitational potential $U^{(0)}_{1}=\triangle^{-1}_{0}\left(\abs{\sigma_{1}^{(0)}}^2\right)$ is monotonically increasing to zero as can be seen in Fig.~1. %\ref{Fig/Tab1}.
It follows from Eq.~(\ref{Eq:SigmaAnsatz}) that linear stability requires that the real part of each eigenvalue $\lambda$ of the system~(\ref{Syst}) with $\ell=1$ is zero. To compute these eigenvalues we used the methodology described in the previous subsection and apply a similar methodology to the equivalent system~(\ref{Eq:LinSyst}) in order to check the validity of our results.

In our previous work~\cite{Roque:2023sjl} we conjectured that under radial perturbations ($J=0$) these configurations are stable and correspond to a local minimum of the conserved energy functional $\mathcal{E}$ when restricted to purely radial perturbations. Our new results are in agreement with this conjecture and indicate that they are also stable under linear nonspherical perturbations with $J=1,2,\dots, 10$. That is, we found only purely oscillatory modes with strictly imaginary eigenvalues that come in pairs $(\lambda, -\lambda)$ as was discussed in subsection~\ref{Sec:IVA}.

Table~\ref{Fig/Tab1} presents the three lowest positive frequencies $\lambda$ for the first seven $J$ values. Notice that for even $J=0, 2$ values the first eigenvalue corresponds to the stationary mode with $\lambda_{st}:=\lambda=0$ discussed in subsection~\ref{Sec:IVB}. Numerically we can \textit{identify} these eigenvalues because although they are not zero to machine precision, they are several orders smaller in magnitude than the remaining eigenvalues. For example, for $J=0$ and $\ell=1$ the ratio with the first non-stationary eigenvalue is $|\lambda_{st}/\lambda|\sim 10^{-3}$. Their eigenfunctions fulfill the relation~Eq.~(\ref{Eq:Rel_Eing_Sta}). In the case of odd values for $J$, the stationary modes belong to the odd-parity sector 
%(e.g., the $L=\ell$ case is part of this sector) 
which we do not study numerically because it only contributes to oscillatory modes.\footnote{In our previous work~\cite{Roque:2023sjl}, in Tables III and V we did not present the stationary eigenvalues because in this case $J=0$ and the zero eigenvalues correspond to infinitesimal rotations in the phase of the unperturbed wave function, as discussed above.}

Finally, we observe from Table~\ref{Fig/Tab1} that (when excluding the stationary modes) the slowest oscillating nonspherical modes with the largest period have total angular momentum $J=1$.

%%%%%
\subsection{Ground states in other $\ell-$boson stars}
%%%%%

Next, we generalize the above study to configurations with $\ell=0,2,3,\dots, 6$ and non-radial linear perturbations with $J=1,\dots, 10$. This extends our previous results presented in Ref.~\cite{Roque:2023sjl}, where it was demonstrated that these configurations are stable under radial perturbations $J=0$. As a check of our results, we computed the respective eigenfunctions $\mathcal{A}_{JM},\mathcal{B}_{JM}$ for every type of eigenvalues $\lambda$ found: real, purely imaginary, complex with nonzero real and imaginary parts, and we validated that these satisfy the properties discussed in Sec.~\ref{Sec:Properties}. In particular, we verified the quadruple symmetry and the fact that $i\lambda(\mathcal{A}_{JM},\mathcal{B}_{JM})_{\text{even}}$ is real.

Similar to the $\ell=1$ case, we show in Table~\ref{Fig/Tab2} the three lowest positive eigenvalues for the configurations $\ell=0,2,3$ with $J=0, 1,2, 3$. As can be appreciated, similar to configurations with $\ell=1$, $(\ell=0)$-boson stars only exhibit purely oscillation modes and a stationary solution for $J=0$. In contrast, for $\ell=2$ and $3$ we found a real eigenvalue in the sector with total angular momentum $J=2$. (Strictly speaking, this eigenvalue has a nonzero small imaginary part; however a  convergence study reveals that by increasing the number $\mathsf{N}$ of Chebyshev points the imaginary part converges to zero.)

Figure~\ref{Fig3} shows the components of the eigenfunctions $\mathcal{A}_{JM}$ (left panel) and $\mathcal{B}_{JM}$ (right panel) corresponding to even-parity modes with $J=2$ and a real eigenvalue, corresponding to the solutions discussed in subsections~\ref{Sec:IVC} and~\ref{Sec:IVD}. In particular, we found that $\mathcal{A}_{JM}$ is purely imaginary, $\mathcal{B}_{JM}$ real, and interestingly, the numerical results indicate that the $L=\ell=2$ component of $B_{JM}{}^L$ seems to be proportional to the background solution, that is, $B_{JM}{}^2\sim\sigma_2^{(0)}$.

Returning to Table~\ref{Fig/Tab2}, we observe that for $\ell = J=3$, complex eigenvalues with nonvanishing real and imaginary parts appear. In fact, we found that this type of eigenvalue is also present in configurations with $2\leq\ell\leq 9$ (see the left panel in Fig.~\ref{Fig4}). The corresponding modes grow exponentially in time implying that the underlying background solution is linearly unstable. This leads us to conjecture that nonrelativistic $\ell-$boson stars with $\ell\geq 3$ possess at least one exponentially in time growing mode characterized by a complex eigenvalue $\lambda$ with $\lambda_I\neq 0$. 

Summarizing, for ground state configurations of the $\ell-$boson stars we verified that the eigenvalues of the linearized system~(\ref{EqGP}) satisfy the properties discussed in Sec.~\ref{Sec:Properties}. Furthermore, we found that they possess the following features:
\begin{itemize}
    \item[(a)] Configurations with $\ell=0, 1$ only present oscillating modes whose largest periods correspond to the smallest $J$ values.
    \item[(b)] A family of stationary modes of the form Eq.~(\ref{Eq:Rel_Eing_Sta}) exist for a given set of values $\ell$, $J\in \{ 0,1,\ldots, 2\ell \}$ and $|M|\leq J$. For $\ell=2,4, \dots, 2n$ with $n\in \mathbb{N}$ and $J\neq 0$, they have an angular dependency that is different from the background solution; hence they are expected to give rise to stationary nonspherical deformation in the nonlinear case.
    \item[(c)] Configurations with $\ell>1$ have in the even-parity sector with $J=2$ a real eigenvalue, for which all components of $\mathcal{B}_{JM}$ ($\mathcal{A}_{JM}$) are real (purely imaginary), and the component $B_{2M}{}^\ell$ is proportional to $\sigma_{\ell}^{(0)}$. These modes are exponentially growing in time. 
    \item[(d)] Configurations with $\ell\geq3$ have at least one unstable mode that grows exponentially in time and is characterized by a complex eigenvalue with nonvanishing real and imaginary parts.
    \item[(e)] Perturbations with large total angular momentum $J$ have only purely oscillatory modes. As can be seen from the left panel of Fig.~\ref{Fig4}, the real parts of the eigenvalues $\lambda$ vanish above a certain value of $J$, leaving only oscillatory modes. This result is compatible with the analytical results of subsection~\ref{Sec:IVE}, where the absence of unstable modes for high enough values of $J$ was proven. Therefore, the lowest $J$ modes are the ones that determine the linear stability of the nonrelativistic $\ell$-boson stars.
\end{itemize}

%%%%%
\subsection{Excited states in nonrelativistic $\ell$ boson stars}
%%%%%
Finally, to close this section, we discuss briefly the mode stability of excited $\ell-$boson stars, i.e. background configurations with $n>0$ nodes.

In our previous article~\cite{Roque:2023sjl} we conjectured that these configurations are linearly unstable -- with exponentially in time growing modes -- under radial perturbations $J=0$. The findings in this section allow us to strengthen this conjecture: in addition to the unstable modes reported in our previous work, here we found unstable non-spherical modes characterized by purely real or complex eigenvalues. Furthermore, we found non-spherical stationary and purely oscillatory modes. Our results support the conclusion that excited states of nonrelativistic $\ell$-boson stars are unstable.

Similar to the ground state configurations, under perturbations with large total angular momentum excited configurations only have oscillatory modes. In the right panel of Fig.~\ref{Fig4} we show the number of eigenvalues with non-zero real parts as a function of $J$. Notice that in contrast to the ground state configurations, the real eigenvalues are not limited to the $J=2$ sector; however they are constrained to even values of $J$.

%%%%%%%%%%%%%%%%%%%%%%%%%%%%%%%%%%%%%%%
\section{Conclusions}
\label{Sec:Conclusions}
%%%%%%%%%%%%%%%%%%%%%%%%%%%%%%%%%%%%%%%

The main result of this paper is the discovery that nonrelativistic $\ell$-boson stars with angular momenta $\ell > 1$, when slightly perturbed from their equilibrium state, are subject to unstable non-radial modes. This includes, in particular, the ground state configurations  which had previously been shown to be linearly stable with respect to radial perturbations~\cite{Roque:2023sjl}. We reached this conclusion by decoupling the linearized $N$-particle Schr\"odinger-Poisson system into a family of radial eigenvalue problems obtained by expanding the linearized wave function in terms of tensor spherical harmonics. While only purely oscillatory modes were found for ground state configurations with $\ell=0$ and $\ell=1$, we found exponentially in time growing modes for ground state and excited configurations with $\ell=2,3,\ldots,9$. These unstable modes have total angular momentum numbers $J$ lying between $1$ and a finite limit depending on $\ell$, and hence they give rise to a non-spherical gravitational potential. This leads us to the conjecture that all $\ell$-boson stars with $\ell > 1$ are unstable with respect to non-spherical linearized perturbations.

Although the configurations with $\ell=2,3,\ldots, 9$ have been found to be unstable, they could still be relevant if they decayed in a very slow fashion (for example, with a timescale larger than the age of the Universe). For this reason, it is important to quantify their lifetimes which we define by $t_{\text{life}} := 1/\lambda_R$, with $\lambda_R$ the real part of the eigenvalue associated with the fastest growing mode. Focusing on the ground state configurations with $\ell = 2, 3$, it turns out that the fastest growing modes are the ones associated with purely real eigenvalues with a total angular momentum $J=2$. Their respective lifetimes are  $t_{\text{life}}\approx 1/0.0064437\;t_c/N^2$ and $t_{\text{life}}\approx 1/0.0090776 \; t_c/N^2$, where the time scale $t_c$ is defined in Eq.~(\ref{Eq:characteristicQ}) and $N$ refers to the total particle number. Accordingly, $t_{\text{life}}$ scales like $1/(N^2\mu^5)$ where $\mu$ is the rest mass of the particles. For the sake of illustration, let us compute the lifetime for two typical astrophysical objects: a dwarf planet with mass of the order of $10^{16} \text{kg}$ and radius $R\approx 200\text{km}$ and a dark matter galactic halo with mass of the order of $10^{10}$ solar masses and radius $R\approx 1 \text{Kpc}$. For both $\ell=2$ and $\ell=3$, these objects can be mimicked by non-relativistic ground state $\ell$-boson stars with $N\approx 10^{55}$ and $N\approx 10^{97}$ bosons of mass $\mu\approx 10^{-3}$ and $\mu\approx 10^{-22} \text{eV}/c^2$, respectively~\cite{Roque:2023sjl}.
%\footnote{ To mimic these astrophysical objects we assume that the $0.99 M$ of the dimensionless total mass $M=N$ is contained at $R_{99}:=r\approx 100$ and that the physical radius of the configuration is much larger than its Schwarzschild radius $R_s^{phys} := 2G M^{phys}/c^2$, which leads to the restriction
%\begin{align}
%    \frac{R_s^{phys}}{R_{99}^{phys}}&=
%\frac{(2N)^2}{R_{99}}\left(\frac{\mu}{m_{\text{pl}}}\right)^4\ll 1,
%\label{PhysicRel}
%\end{align}
%where $m_{\text{pl}} = \sqrt{\hbar c/G} = 2.17643\times10^{-8}$kg is the Planck mass. The restriction~(\ref{PhysicRel}) guarantees that we are in the non-relativistic regime. In the case that $R_s^{phys}$ is comparable to $R_{99}$ one needs to consider the relativistic $\ell$-boson stars~\cite{Alcubierre:2018ahf}.} 
The resulting lifetimes are of the order of $3\text{hr}$ for the dwarf planet analogue and of $10^{6}\text{yr}$ for the galactic halo model, much smaller than the typical lifetimes associated with these objects.

%In conclusion, as a consequence of the unstable non-radial modes, the non-relativistic $\ell-$boson stars with angular momentum $\ell>1$ have lifetimes that are generally too short to be considered stable for practical purposes.

In addition to the unstable modes, our analysis also revealed the existence of nonspherical stationary solutions of the linearized system for each $\ell > 0$ configuration. As stated previously, these modes indicate the bifurcation of new branches of nonspherical stationary deformations of the $\ell$-boson stars, and it should be interesting to establish their existence and analyze their properties.

The methodology developed in this article for analyzing the linearized system should also be applicable to more general boson star configurations, including multistate~\cite{Bernal:2009zy, Urena-Lopez:2010zva} and multi-$\ell$ multistate  configurations~\cite{Alcubierre:2022rgp} in their nonrelativistic limit. For instance, it would be interesting to analyze whether a ground state $\ell=2$-boson star can be stabilized by adding an $\ell=0$ field to it.

We expect the non-radial instabilities found in this article to carry over to the fully relativistic $\ell$-boson stars~\cite{Alcubierre:2018ahf} with $\ell > 1$.

%%%%%%%%%%%%%%%%%%%%%%%%%%%%%%%%%%%%%%%
\subsection*{Acknowledgements}
%%%%%%%%%%%%%%%%%%%%%%%%%%%%%%%%%%%%%%%

It is a pleasure to thank Argelia Bernal, Alberto Diez-Tejedor, and Emilio Tejeda for enlightening discussions. This work was partially supported by CONAHCyT Network Projects No.~376127 ``Sombras, lentes y ondas gravitatorias generadas por objetos compactos astrof\'isicos'', by a CIC grant to Universidad Michoacana de San Nicol\'as de Hidalgo, and CONAHCyT-SNI. A.A.R. also acknowledges funding from a postdoctoral fellowship from ``Estancias Posdoctorales por M\'exico para la Formaci\'on y Consolidaci\'on de las y los Investigadores por M\'exico''. E.C.N. was supported by a CONAHCyT doctoral scholarship.

%%%%%%%%%%%%%%%%%%%%%%%%%%%%%%%%%%%%%%%
\appendix 
%%%%%%%%%%%%%%%%%%%%%%%%%%%%%%%%%%%%%%%

%%%%%%%%%%%%%%%%%%%%%%%%%%%%%%%%%%%%%%%
\section{Tensor spherical harmonics}
\label{App:SphericalTensorHarmonics}
%%%%%%%%%%%%%%%%%%%%%%%%%%%%%%%%%%%%%%%

In this appendix we recall the definition of the tensor spherical harmonics (TSH) and briefly review a few basic known facts about them which are relevant for this article. A more extended discussion and additional properties can be found in Ref.~\cite{Khersonskii:1988krb}. 

TSH describe the angular distribution and polarization of spin $S$ particles with total angular momentum $J$, total magnetic quantum number $M$, and orbital angular momentum $L$. To define them, consider the class $V := L^2(\Real^3,\Complex^{2S+1})$ of wave functions $\Psi: \Real^3\to \Complex^{2S+1}$ which are Lebesgue square-integrable. The rotation group $SO(3)$ induces a unitary representation $U(R): V\to V$ defined by
\begin{equation}
(U(R)\Psi)(\vec{x}) := D^{(S)}(R)\Psi(R^{-1}\vec{x}),\qquad
\vec{x}\in \Real^3,
\end{equation}
for $\Psi\in V$ and $R\in SO(3)$, where $D^{(S)}(R): \Complex^{2S+1}\to\Complex^{2S+1}$ is a unitary representation of $SO(3)$ on $\Complex^{2S+1}$. The corresponding representation of the Lie algebra leads to the total angular momentum operator $\hat{\vec J}$ (i.e. the generators associated with rotations along the coordinate axes divided by $i$)
\begin{equation}
\hat{\vec J} = \hat{\vec L} + \hat{\vec S},
\end{equation}
with $\hat{\vec L} := -i\vec{x}\wedge\vec\nabla$ the orbital angular momentum operator and $\hat{\vec S}$ the spin operator. Since the components of $\hat{\vec L}$ and $\hat{\vec S}$ commute with each other, one can check that the following operators commute among themselves: $\hat{J}^2$, $\hat{J}_z$, $\hat{L}^2$, $\hat{S}^2$. The TSH are particular wave functions $Y^{JM}{}_{L S}\in V$ which are eigenfunctions of these operators. They are constructed from the standard (scalar) spherical harmonics $Y^{Lm}$ (which are eigenfunctions of the operators $\hat{L}^2$ and $\hat{L}_z$) and basis spin functions $\xi^{S \sigma}$ (which are eigenfunctions of $\hat{S}^2$ and $\hat{S}_z$) in accordance with the addition of angular momenta in quantum mechanics:
\begin{equation}
\label{TensorSphericalHar}
Y^{JM}{}_{L S}(\vartheta,\varphi) := \sum\limits_{m,\sigma}
C^{JM}{}_{LmS\sigma} Y^{Lm}(\vartheta,\varphi)\xi^{S \sigma},
\end{equation}
Here, $C^{JM}{}_{LmS\sigma}$ denote the Clebsch-Gordan coefficients and $J$, $S$ are nonnegative integer or half-integer numbers. Given a pair $(J,S)$, the admissible values for $L$ and $M$ are: $L = |J - S|, |J - S| + 1, \dots, J + S$ and $M = -J, \ldots ,J$. The basis spin functions $\xi^{S\sigma}$ satisfy the conditions
\begin{equation}
    \hat{S}^2 \xi^{S\sigma} = S(S +1)\xi^{S\sigma}, \quad \hat{S}_z \xi^{S\sigma} = \sigma \xi^{S\sigma},\quad
\sigma = -S,\ldots,S,
\end{equation}
and since $\hat{S}_z$ is self-adjoint, they form an orthonormal basis of $\Complex^{2S+1}$ after suitable normalization.

%The basis spin functions are functions of a discrete argument $\mu = -S, -S + 1, \ldots, S$ which is the projection on $z$ axis of the spin. This functions can be seen as a vector with $2S+ 1$ components

%\begin{align}
%    \xi^{S\sigma} &= \left(\xi^{S\sigma}(-S), \xi^{S\sigma}(-S + 1, \ldots, S)\right)^T,\\
%    [\xi^{S\sigma}]^* &= \left(\xi^{S\sigma}(-S), \xi^{S\sigma}(-S + 1, \ldots, S)\right),
%\end{align}
%with the components given by $\xi^{S\sigma}(\mu) = \delta_{\sigma\mu}$. Furthermore, the basis spin functions satisfies the orthonormality and completeness conditions

%\begin{align}
%  [\xi^{S\sigma}]^* \xi^{S\sigma'} = \delta_{\sigma\sigma'}, \quad \sum_{\sigma = -S}^{S} \xi^{S\sigma} [\xi^{S\sigma}]^* = \hat{I},
%\end{align}
%which implies that $\xi^{S\sigma}$ constitutes a complete orthonormal set of eigenfunctions. 

Using these properties and those of the scalar spherical harmonics, it is not difficult to verify that
\begin{align}
    \hat{J}^2 Y^{JM}{}_{L S} &= J(J + 1) Y^{JM}{}_{L S},\\
    \hat{J}_z Y^{JM}{}_{L S} &= M Y^{JM}{}_{L S},\\
    \hat{L}^2 Y^{JM}{}_{L S} &= L(L + 1) Y^{JM}{}_{L S},\\
    \hat{S}^2 Y^{JM}{}_{L S} &= S(S + 1) Y^{JM}{}_{L S}.
\end{align}
Furthermore, it follows that the collection of TSHs with the same $S$ and all possible $J,L,M$ constitutes a complete orthonormal set in the space of Lebesgue square-integrable functions $\mathbb{S}^2\to\Complex^{2S+1}$ on the two-sphere $\mathbb{S}^2$. The orthonormality condition is
\begin{equation}
\int\limits_{\mathbb{S}^2} [Y^{JM}{}_{L S}(\vartheta, \varphi)]^* Y^{JM}{}_{L S}(\vartheta, \varphi) d\Omega = \delta_{JJ'}\delta_{MM'}\delta_{LL'}.
\end{equation}
%and the completeness of these functions is given by
%\begin{align}
%    &\sum_{JLM} [Y^{JM}{}_{L S}(\vartheta, \varphi)]^\mu [Y^{JM}{}_{L S}(\vartheta', \varphi')]^{\nu*} = \delta_{\mu\nu} \times\\
%    &\times \delta(\cos{\vartheta} - \cos{\vartheta'}) \delta(\varphi - \varphi'),
%\end{align}
%where $[Y^{JM}{}_{L S}(\vartheta, \varphi)]^\mu$ and $[Y^{JM}{}_{L S}(\vartheta, \varphi)]^\nu$ ($\mu, \nu = -S, -S + 1, \ldots, S$) denotes the $\mu$ and $\nu$ components of $Y^{JM}{}_{L S}(\vartheta, \varphi)$ respectively. 
%
When $S = \ell$ is an integer, one can choose the representation $D^{(S)}(R)$ to be described by real-valued matrices, which implies that $\hat{\vec S}$ is purely imaginary. Hence, $\overline{\hat{S}_z\xi^{S\sigma}} = -\hat{S}_z\overline{\xi^{S\sigma}}$ which implies that the basis spin functions can be chosen such that they satisfy
\begin{equation}
\overline{\xi^{\ell \sigma}} = (-1)^\sigma \xi^{\ell -\sigma},\qquad
\sigma = -\ell,-\ell+1,\ldots,\ell.
\end{equation}
Together with the corresponding relation $\overline{Y^{\ell m}} = (-1)^m Y^{\ell -m}$ for the scalar spherical harmonics and the identity 
\begin{equation}
C^{JM}{}{}_{Lm \ell \sigma} = (-1)^{L + \ell + J} C^{J-M}{}{}_{L -m \ell -\sigma},
\end{equation}
for the Clebsch-Gordan coefficients one obtains the useful relation
\begin{equation}
\overline{Y^{JM}{}_{L\ell}} = (-1)^{J+M+L+\ell} Y^{J -M}{}_{L\ell}.
\label{ComplexConjugationTensorSphericalBis}
\end{equation}
between the TSH and its complex conjugate.

%for instance, the following explicit expression for the basis spin functions
%
%\begin{equation}
%\xi^{\ell\sigma} := \frac{1}{\sqrt{2}}
%\left\{ \begin{array}{rl}
% e_\sigma + i(-1)^\sigma e_{-\sigma}, & \sigma=-\ell,\ldots,-1,\\
% \sqrt{2} e_0, & \sigma = 0,\\
% - i e_\sigma + (-1)^\sigma e_{-\sigma}, & \sigma = 1,\ldots,\ell,
%\end{array} \right.
%\end{equation}
%
%with $e_\sigma$ denoting the vector in $\Complex^{2\ell+1}$ which has a one in its $\sigma$'s entry and zeros in its other entries (assuming the components of vectors in $\Complex^{2\ell+1}$ run from $-\ell$ to $\ell$). 

As an example, consider $S=\ell=1$, in which case one can choose $D^{(1)}(R) = R$ such that
\begin{equation}
\hat{S}^2 = 2\left(\begin{array}{rrr}
1 & 0 & 0 \\
0 & 1 & 0 \\
0 & 0 & 1
\end{array} \right),
\quad
\hat{S}_z = \frac{1}{i}
\left( \begin{array}{rrr}
0 & 1 & 0 \\
-1 & 0 & 0 \\
0 & 0 & 0
\end{array} \right).
\end{equation}
In this case the basis spin functions can be chosen as
\begin{equation}
\xi^{11} := -\frac{1}{\sqrt{2}}\left( \hat{e}_x + i\hat{e}_y \right),\quad
\xi^{10} := \hat{e}_z,
\end{equation}
and $\xi^{1-1} := -\overline{\xi^{11}}$.

Finally, we summarize some useful expressions for the Clebsch-Gordan coefficients which are used throughout this article. First, when $J=0$, one has the simple expression
\begin{equation}
\label{JZeroCB}
    C^{00}{}_{\ell m \ell \sigma} = \frac{(-1)^{\ell - m}}{\sqrt{2\ell + 1}} \delta_{m, -\sigma}.
\end{equation}
For $J=\ell=1$ and $M=0$ one has
\begin{align}
C^{10}{}_{1010} = 0,\quad
C^{10}{}_{1-111} = C^{10}{}_{111-1} = \frac{1}{\sqrt{2}}.
\end{align}

%%%%%%%%%%%%%%%%%%%%%%%%%%%%%%%%%%%%%%%
\section{Explicit form of the perturbation equations for nonrelativistic $\ell$-boson stars with $\ell=0, 1, 2$}
\label{App:ExplicitLinSys}
%%%%%%%%%%%%%%%%%%%%%%%%%%%%%%%%%%%%%%%

This appendix presents special cases of the linearized problem~(\ref{EqGP}) in a more explicit way for the particular cases $\ell=0,1,2$. Recall that this system is given by
\begin{subequations}\label{ApC:eqPrinc}
\begin{align}
    i\lambda A_{JM}{}^L &=\left(\hat{\mathcal{H}}_{L}^{(0)}-E\right)B_{JM}{}^L,\\
    i\lambda B_{JM}{}^L &=\left(\hat{\mathcal{H}}_{L}^{(0)}-E\right)A_{JM}{}^L + 2Q_{JM}{}^L,
\end{align}
\end{subequations}
where the fields $(A_{JM}{}^L, B_{JM}{}^L)$ are labeled by the numbers $(JML)$ such that $J = 0, 1, 2, \ldots$, $|M| \leq J$ and $|J-\ell|\leq L\leq J+\ell$ and the operator $\mathcal{\hat H}_{L}^{(0)}$ was defined in Eq.~(\ref{Eq:HL}). The function $Q_{JM}{}^L(r)$ is defined as (see Eq.~(\ref{Eq:QJML}))
\begin{eqnarray}
Q_{JM}{}^L(r) &=& \sigma_\ell^{(0)}(r)\sum\limits_{L'=|J-\ell|}^{J+\ell}
\frac{\sqrt{(2L+1)(2L'+1)}}{2J+1} 
\nonumber\\
 &\times& C^{J0}{}_{L0\ell 0} C^{J0}{}_{L'0\ell 0} 
 \Delta_J^{-1}\left(\sigma_\ell^{(0)} A_{JM}{}^{L'} \right)(r),
\qquad
\end{eqnarray}
where the Clebsch-Gordan coefficients $C^{J 0}{}_{L0\ell 0}$ can be computed using the explicit formula (32) in Sec. 8.5.2 of Ref~\cite{Khersonskii:1988krb} and where $\triangle_J^{-1}$ was defined in Eq.~(\ref{Eq:LapJInv}).\\

\textbf{Radial perturbations:} For the particular value $J=0$, we have that $L=\ell$ and the system~(\ref{ApC:eqPrinc}) reduces to the system (26) in Ref.~\cite{Roque:2023sjl}:
\begin{subequations}\label{ApC:eq:J=0}
{\normalsize %\fontsize{9.5pt}{10pt}
  \setlength{\abovedisplayskip}{6pt}
  \setlength{\belowdisplayskip}{\abovedisplayskip}
  \setlength{\abovedisplayshortskip}{0pt}
  \setlength{\belowdisplayshortskip}{3pt}
\begin{align}
    i\lambda A_{0 0}{}^{\ell} &=\left(\hat{\mathcal{H}}_{\ell}^{(0)}-E\right)B_{0 0}{}^{\ell},\\
    i\lambda B_{0 0}{}^{\ell}&=\left(\hat{\mathcal{H}}_{\ell}^{(0)}-E\right)A_{0 0}{}^{\ell} + 2\sigma_{\ell}^{(0)} \Delta_{0}^{-1}\left (\sigma_{\ell}^{(0)}{A_{0 0}{}^{\ell}}\right),
\end{align}
}%
\end{subequations}
where we used the identity $C^{0 0}{}_{\ell 0 \ell 0}=(-1)^{\ell}/\sqrt{2\ell+1}$ obtained from Eq.~(\ref{JZeroCB}).

\textbf{Nonrelativistic ($\ell=0$)-boson star:} In this case we have $J = 0, 1, 2, \ldots$, $|M|\leq J$ and $L = J$. Using the fact that $C^{J 0}{}_{J 0 0 0}=1$ the system~(\ref{ApC:eqPrinc}) reduces to 
%the system~(\ref{ApC:eq:J=0}) with the following changes:  $A_{0 0}{}^{\ell}\to A_{J M}{}^{J}$, $B_{0 0}{}^{\ell}\to B_{J M}{}^{J}$, $\hat{\mathcal{H}}_{\ell}^{(0)}\to \hat{\mathcal{H}}_{J}^{(0)}$, $\Delta_{0}^{-1}\to \Delta_{J}^{-1}$, and $\sigma_{\ell}^{(0)}\to \sigma_{0}^{(0)}$:
%
\begin{subequations}\label{ApC:eq:ell=0}
\begin{align}
    i\lambda A_{J M}{}^{J} &=\left(\hat{\mathcal{H}}_{J}^{(0)}-E\right)B_{J M}{}^{J},\\
    i\lambda B_{J M}{}^{J}&=\left(\hat{\mathcal{H}}_{J}^{(0)}-E\right)A_{J M}{}^{J} + 2\sigma_{0}^{(0)} \Delta_{J}^{-1}\left (\sigma_{0}^{(0)}{A_{J M}{}^{J}}\right),
\end{align}
\end{subequations}
which provides the relevant equations describing nonspherical linearized perturbations of the standard nonrelativistic boson stars.\\

\textbf{Nonrelativistic ($\ell=1$)-boson star:} In this situation we have two possible cases depending on the value of $J$. In the first case, when $J < \ell$, i.e., $J=0$, one must have $L=1$ corresponding to the system~(\ref{ApC:eq:J=0}) with $\ell=1$. The other case represents perturbations with $J\geq\ell$ which have $L \in \{J-1,J, J+1\}$. Using that
\begin{subequations}
\begin{align}
{C^{J0}}_{(J-1)010} &= \sqrt{\frac{J}{2J-1}},\\
{C^{J0}}_{(J+1)010} &= -\sqrt{\frac{J+1}{2J+3}},
\end{align}
\end{subequations}
and ${C^{J0}}_{J010}=0$, we arrive at the system~(\ref{Eqell1}) with $\alpha_{JM} := \big(A_{J M}{}^{J-1}, A_{J M}{}^{J+1}\big)^{T}$ and $\beta_{JM} := \big(B_{J M}{}^{J-1}, B_{J M}{}^{J+1}\big)^{T}$ and the system~(\ref{Eq:LinSyst2}) replacing $(A_{JM}^{(2)},B_{JM}^{(2)})$ with $\big(A_{J M}{}^{J}, B_{J M}{}^{J}\big)$.\\

\textbf{Nonrelativistic ($\ell=2$)-boson star:} Similar to the previous case, there are two possibilities: $J<\ell$, i.e., $J\in \{0, 1\}$, and $J\geq\ell$. For $J=0$, we have that $L=2$ corresponding to the system~(\ref{ApC:eq:J=0}) with $\ell=2$. The value $J=1$ implies that $L\in\{1,2,3\}$ and the perturbations are determined by the system~%(\ref{ApC:eq:ell=1}) with the following substitutions: $\sigma_1^{(0)}\mapsto \sigma_2^{(0)}$, $\alpha_{JM}\mapsto \alpha_{1M} := \big(A_{1 M}{}^{J-1}, A_{1 M}{}^{J+1}\big)^{T}$, $\xi_{JM} \mapsto \xi_{1M} := \big(A_{1 M}{}^{J}, B_{1 M}{}^{J}\big)^{T}$, and $\beta_{JM}\mapsto \beta_{1M} := \big(B_{1 M}{}^{J-1}, B_{1 M}{}^{J+1}\big)^{T}$ and $J=2$.

\begin{widetext}
\begin{subequations}\label{ApC:eq:ell=1}
\begin{align}
i\lambda\alpha_{1M} &= 
\left( \begin{array}{cc}
\mathcal{\hat H}_{1}^{(0)} - E & 0 \\
0 & \mathcal{\hat H}_{3}^{(0)} - E
\end{array} \right)\beta_{1M},\label{ell1_a}\\
i\lambda\beta_{1M} &= 
\left( \begin{array}{cc}
\mathcal{\hat H}_{1}^{(0)} - E & 0 \\
0 & \mathcal{\hat H}_{3}^{(0)} - E
\end{array} \right)\alpha_{1M}
 + \frac{2\sigma_2^{(0)}}{5}
\left(\begin{array}{cc} 2 & -\sqrt{6} \\ -\sqrt{6} & 3 \end{array} \right)
 \Delta_1^{-1}\left ( \sigma_2^{(0)}\alpha_{1M} \right),\label{ell1_c}\\
 i\lambda\gamma_{1M} &= 
\left( \begin{array}{cc}
0 & \mathcal{\hat H}_{2}^{(0)} - E \\
\mathcal{\hat H}_{2}^{(0)} - E & 0
\end{array} \right)\gamma_{1M},\label{ell1_b}
\end{align}
\end{subequations}
\end{widetext}
where $\alpha_{1M} := \big(A_{1 M}{}^{1}, A_{1 M}{}^{3}\big)^{T}$, $\beta_{1M} := \big(B_{1 M}{}^{1}, B_{1 M}{}^{3}\big)^{T}$, $\gamma_{1M} := \big(A_{1 M}{}^{2}, B_{1 M}{}^{2}\big)^{T}$, and $M = -1, 0, 1$.

Finally, the linear perturbations with $J\geq\ell$ i.e., $J \in \{2, 3,  \ldots$\},  $L \in \{J-2, \ldots, J+2\}$ are described by the system
\begin{widetext}
\begin{subequations}\label{ApC:eq:ell=2}
\begin{align}
i\lambda\tilde\alpha_{JM} &= 
\left( \begin{array}{ccc}
\mathcal{\hat H}_{J-2}^{(0)} - E & 0 & 0\\
0 & \mathcal{\hat H}_{J}^{(0)} - E & 0 \\
0 & 0 & \mathcal{\hat H}_{J+2}^{(0)} - E
\end{array} \right) \tilde\beta_{JM},\label{ell2_a}\\
i\lambda\tilde\beta_{JM} &= 
\left( \begin{array}{ccc}
\mathcal{\hat H}_{J-2}^{(0)} - E & 0 & 0 \\
0& \mathcal{\hat H}_{J}^{(0)} - E & 0 \\
0 & 0& \mathcal{\hat H}_{J+2}^{(0)} - E
\end{array} \right)\tilde\alpha_{JM}
 + \frac{2\sigma_2^{(0)}}{2J-1}
\left(\begin{array}{ccc} \mathcal{J}_{1} \mathcal{J}_{2}/\mathcal{J}_{3} & -\mathcal{J}_{1} & \mathcal{J}_{2} \\ -\mathcal{J}_{1} & \mathcal{J}_{1}\mathcal{J}_{3}/\mathcal{J}_{2}& -\mathcal{J}_{3}\\
\mathcal{J}_{2} & -\mathcal{J}_{3} & \mathcal{J}_{3} \mathcal{J}_{2}/\mathcal{J}_{1}\end{array} \right)
 \Delta_J^{-1}\left (\sigma_2^{(0)}\tilde\alpha_{JM} \right),\label{ell2_c}\\
i\lambda\tilde\gamma^{\pm}_{JM} &= 
\left( \begin{array}{cc}
0 & \mathcal{\hat H}_{J\pm1}^{(0)} - E \\
\mathcal{\hat H}_{J\pm1}^{(0)} - E & 0
\end{array} \right)\tilde\gamma^{\pm}_{JM},\label{ell2_b},
\end{align}
\end{subequations}
\end{widetext}
where now 
\begin{subequations}
\begin{align}
\tilde\alpha_{JM} &:= \big(A_{J M}{}^{J-2}, A_{J M}{}^{J},A_{J M}{}^{J+2}\big)^{T},\\
\tilde\beta_{JM} &:= \big(B_{J M}{}^{J-2}, B_{JM}{}^{J}, B_{JM}{}^{J+2}\big)^{T},\\
\tilde\gamma^{\pm}_{JM} &:= \big(A_{J M}{}^{J\pm1}, B_{J M}{}^{J\pm1}\big)^{T},
\end{align}
\end{subequations}
and
\begin{subequations}
\begin{align}
\mathcal{J}_{1}&:=\sqrt{\frac{3J^2(J^2-1)}{2(2J+1)(2J+3)}},\\
\mathcal{J}_{2}&:=\frac{3}{2(2J+1)}\sqrt{\frac{J(J^2-1)(J+2)(2J-1)}{2J+3}},\\
\mathcal{J}_{3}&:=\frac{J+1}{2J+3}\sqrt{\frac{3J(J+2)(2J-1)}{2(2J+1)}}.
\end{align}
\end{subequations}

%%%%%%%%%%%%%%%%%%%%%%%%%%%%%%%%%%%%%%%
\section{Decoupling the perturbed evolution equation}
\label{App:Decoupling}
%%%%%%%%%%%%%%%%%%%%%%%%%%%%%%%%%%%%%%%

In this appendix we show that the evolution equation~(\ref{Eq:PerturbationEq}) for the linearized field $\chi$ can be decoupled by expanding $\chi$ in terms of tensor spherical harmonics. Writing
%From the perturbed evolution equation~(\ref{Eq:PerturbationEq})
%\begin{align}    
%i\frac{\partial\chi}{\partial t} = \bigg(\hat{\mathcal{H}}_0-E\bigg)\chi + 2\triangle^{-1}\bigg(\Re{\chi_0^{*}\chi}\bigg)\chi_0,
%\end{align}
%we can expand the fields $\chi_0, \chi$ in terms of tensor spherical harmonics ${Y^{JM}}_{L\ell}$ as
\begin{subequations}
\begin{align}
    \chi_0&:= \sqrt{4\pi}\sigma_\ell^{(0)}(r){Y^{00}}_{\ell\ell}(\vartheta,\varphi),\\
    \chi&:=\sum_{JLM} {X_{JM}}^{L}(t, r) {Y^{JM}}_{L\ell}(\vartheta,\varphi),
\end{align}
\end{subequations}
Eq.~(\ref{Eq:PerturbationEq}) yields
\begin{align}    
i\frac{\partial}{\partial t}{X_{JM}}^{L}(t, r)  = \bigg(\hat{\mathcal{H}}_L^{(0)} - E\bigg){X_{JM}}^{L}(t, r)  + {q_{JM}}^{L}(t,r),\label{EqAnexEvol}
\end{align}
where
\begin{align}
    q_{JM}{}^L(t, r) &= \sigma_\ell^{(0)}(r)\sum\limits_{L'=|J-\ell|}^{J+\ell}
\frac{\sqrt{(2L+1)(2L'+1)}}{2J+1} 
\nonumber\\
 &\times C^{J0}{}_{L0\ell 0} C^{J0}{}_{L'0\ell 0} 
 \Delta_J^{-1}\left(\sigma_\ell^{(0)}(r) Z_{JM}{}^{L'} (t, r) \right),
\end{align}
and where we have defined $Z_{JM}{}^{L'}(t,r) :={X_{JM}}^{L'}(t, r)+(-1)^{M}\overline{ {X_{J-M}}^{L'}(t, r)}$.

The properties of the Clebsch-Gordan coefficients discussed below Eq.~(\ref{Eq:QJML}) imply that this system further decouples into two subsystems:
\begin{enumerate}
    \item The even-parity sector which contains the values $L = |J-\ell|,|J-\ell|+2,\ldots J+\ell$ and has non-trivial coefficients $q_{JM}{}^L$.
    \item The odd-parity sector which has $L = |J-\ell|+1,|J-\ell|+3,\ldots,J+\ell-1$, for which  $q_{JM}{}^L$ vanishes.
\end{enumerate}
An important consequence of these observations is that in the odd-parity sector the right-hand side of Eq.~(\ref{EqAnexEvol}) is characterized by the self-adjoint operators $\hat{\mathcal{H}}_L^{(0)} - E$ implying a unitary evolution. Consequently, one can have only oscillatory modes in the odd-parity sector, and unstable modes can only arise in the even-parity sector.

%%%%%%%%%%%%%%%%%%%%%%%%%%%%%%%%%%%%%%%
\section{Properties of the second variation of the energy functional}
\label{App:ModeDecompEnergyFunc}
%%%%%%%%%%%%%%%%%%%%%%%%%%%%%%%%%%%%%%%

In this appendix we show two important properties which are satisfied by the second variation $\delta^2\mathcal{E}[\delta u]$ of the conserved energy functional $\mathcal{E}$. The first one is that when $\delta u(\vec{x})$ is replaced with a solution $\chi(t,\vec{x})$ of the linearized equations, $\delta^2\mathcal{E}[\chi]$ is independent of time. This property is indeed expected to hold since it should be inherited from the full nonlinear functional $\mathcal{E}$, which is preserved under the time evolution. The second property consists in the fact that $\delta^2\mathcal{E}$ can be written as a sum over the contributions from each $JM$ mode, when performing the decomposition into tensor spherical harmonics. Of course, this can also be anticipated from the fact that the background is invariant with respect to the total angular momentum operator.

Evaluating the second variation of the energy functional $\delta^2 \mathcal{E}$ given by (\ref{Eq:SecondVariationchi0}) at $\chi$, we get 
\begin{equation}
    \delta^2 \mathcal{E} [\chi] = \left(\chi, [\hat{\mathcal{H}}_0 -E]\chi\right) - 2D[\delta n, \delta n], \quad \delta n = 2\Re{\chi_0^T \chi}.
    \nonumber
\end{equation}
Substituting the ansatz (\ref{Eq:SigmaAnsatz}) for $\chi$ we obtain
\begin{equation}
    \delta^2 \mathcal{E} [\chi] = F\left[\mathcal{A}, \mathcal{B}\right] e^{2\lambda t} + G\left[\mathcal{A}, \mathcal{B}\right] e^{2\lambda^* t} + K\left[\mathcal{A}, \mathcal{B}\right] e^{2\lambda_R t},
\end{equation}
where the functionals $F,G,K$ are defined by
\begin{subequations}
\begin{align}
    F[\mathcal{A}, \mathcal{B}] &:= \left(\overline{\mathcal{A} - \mathcal{B}}, [\hat{\mathcal{H}}_0 - E](\mathcal{A} + \mathcal{B})\right) \\
    \nonumber
    &+ 2\left(\chi_0^T \overline{\mathcal{A}}, \triangle^{-1}(\chi_0^T \mathcal{A})\right), \\
    G[\mathcal{A}, \mathcal{B}] &:= \left(\mathcal{A} + \mathcal{B}, [\hat{\mathcal{H}}_0 - E](\overline{\mathcal{A} - \mathcal{B}})\right) \\ 
    \nonumber
    &+ 2\left(\chi_0^T \mathcal{A}, \triangle^{-1}(\chi_0^T \overline{\mathcal{A}})\right), \\
    K[\mathcal{A}, \mathcal{B}] &:= \left(\mathcal{A} + \mathcal{B}, [\hat{\mathcal{H}}_0 - E](\mathcal{A} + \mathcal{B})\right)\\
    \nonumber
    +& \left(\mathcal{A} - \mathcal{B}, [\hat{\mathcal{H}}_0 - E](\mathcal{A} - \mathcal{B})\right)\\ \nonumber
    &+ 4\left(\chi_0^T \mathcal{A}, \triangle^{-1}(\chi_0^T \mathcal{A})\right).
\end{align}
\end{subequations}
Using the fact that $\hat{\mathcal{H}}_0 - E$ is a self-adjoint operator, it easy to prove $\overline{F[\mathcal{A}, \mathcal{B}]} = G[\mathcal{A}, \mathcal{B}]$. On the other hand, using Eqs.~(\ref{EquationAB}) we also get
\begin{subequations}
\begin{align}
   F[\mathcal{A}, \mathcal{B}] &= i\lambda \left(\overline{\mathcal{A} - \mathcal{B}}, \mathcal{A} + \mathcal{B}\right) + 2\left(\overline{B}, \triangle^{-1}(\chi_0^T \mathcal{A})\chi_0\right),\\ 
   G[\mathcal{A}, \mathcal{B}] &= i\lambda^* \left(\mathcal{A} + \mathcal{B}, \overline{\mathcal{A} - \mathcal{B}}\right) - 2\left(B, \triangle^{-1}(\chi_0^T \overline{\mathcal{A}})\chi_0\right),\\
   K[\mathcal{A}, \mathcal{B}] &= 4 i\lambda \Re\left(\mathcal{A}, \mathcal{B}\right).
\end{align}
\end{subequations}
We see from this that $\overline{F[\mathcal{A}, \mathcal{B}]} = - G[\mathcal{A}, \mathcal{B}]$ which implies that $F = G = 0$. Therefore, we obtain
\begin{equation}
    \delta^2\mathcal{E}[\chi] = 4 i\lambda e^{2\lambda_R t} \Re(\mathcal{A}, \mathcal{B}).
\end{equation}
Let us analyze the implications of this result for the same cases (i) -- (iv) as in subsection~\ref{Sec:IVC}:
\begin{enumerate}
\item[(i)] $\lambda=0$: In this case the second variation of the energy functional is zero and hence trivially time-independent.

\item[(ii)] $\lambda_R > 0$ and $\lambda_I = 0$. In this case Eq.~(\ref{Eq:ProdAB}) implies that $i \lambda (\mathcal{A}, \mathcal{B}) \in \Real$ such that $\Re(\mathcal{A}, \mathcal{B}) = 0$. Again, it follows that $\delta^2\mathcal{E}[\chi] = 0$.

\item[(iii)] $\lambda _R = 0$ and $\lambda_I > 0$. Choosing $\mathcal{A}$ and $\mathcal{B}$ real, one obtains
\begin{equation}
\delta^2 \mathcal{E} [\chi] = -4\lambda_I (\mathcal{A}, \mathcal{B}),
\end{equation}
which is again independent of $t$ and should be compared with Eq.~(\ref{Eq:ProdAB}).

\item [(iv)] $\lambda_R > 0$ and $\lambda_I > 0$. Recall that in this case
$(\mathcal{A}, \mathcal{B}) = 0$, which implies $\delta^2\mathcal{E}[\chi] = 0$.
\end{enumerate}
Summarizing, we conclude that the second variation of the energy functional is indeed time-independent for any solution $\chi(t,\vec{x})$ of the form~(\ref{Eq:SigmaAnsatz}) of the linearized equations. Furthermore, $\delta^2\mathcal{E}[\chi] = 0$ except for case (iii) corresponding to the purely oscillatory modes.

Next, we compute the mode decomposition of the expression
$$
\delta^2\mathcal{E}[\mathcal{A}_R] + \delta^2\mathcal{E}[\mathcal{A}_I] 
 = (\mathcal{A},(\hat{\mathcal{H}}_0 - E)\mathcal{A})
 + 2(\chi_0^T\mathcal{A},\Delta^{-1}[\chi_0^T\mathcal{A}])
$$
appearing on the right-hand side of Eq.~(\ref{Eq:ProdAB}). We focus on the even-parity sector since, as shown in Appendix~\ref{App:Decoupling}, there are no instabilities in the odd-parity sector. Using Eqs.~(\ref{Eq:LapJInv}, \ref{Eq:Q}, \ref{Eq:QJML}) one obtains
\begin{equation}
\delta^2\mathcal{E}[\mathcal{A}_R] + \delta^2\mathcal{E}[\mathcal{A}_I] = \sum\limits_{JM}
\delta^2\mathcal{E}_{JM,even}[\mathcal{A}_{JM}]
\end{equation}
with
\begin{align}
& \delta^2\mathcal{E}_{JM,even}[\mathcal{A}_{JM}] 
\nonumber\\
&= \sum\limits_{\substack{L=|J-\ell| \\ \text{ $J+\ell-L$ even}}}^{J+\ell} \int\limits_0^\infty (A_{JM}{}^L)^*(r)
(\hat{\mathcal H}_L - E)A_{JM}{}^L(r) r^2 dr
\nonumber\\
 &- \frac{1}{2J + 1} \int_{0}^{\infty}\int_{0}^{\infty} \frac{r_<^{J}}{r_>^{J + 1}} a_{JM}(r)^* a_{JM}(\Tilde{r}) r^2 \Tilde{r}^2 dr d\Tilde{r},
\end{align}
where
\begin{equation}
a_{JM}(r) := \sigma^{(0)}_{\ell}(r) \sum\limits_L \sqrt{\frac{2L + 1}{2J + 1}} C^{J0}{}_{L0\ell 0} A_{JM}{}^{L}(r).
\end{equation}
This shows that the second variation of the energy functional can indeed be decomposed in the $JM$ modes.

%%%%%%%%%%%%%%%%%%%%%%%%%%%%%%%%%%%%%%%
\section{Key estimate for the second variation of the energy functional}
\label{App:EstimateSecondVar}
%%%%%%%%%%%%%%%%%%%%%%%%%%%%%%%%%%%%%%%

In this appendix we prove the estimate~(\ref{Eq:EstimateSecVar}) used in Sec.~\ref{Sec:IVE} to rule out the existence of unstable modes with high angular momenta. For this, recall that the second variation of the functional $\mathcal{E}$ is giving by 
\begin{equation}
    \delta^2\mathcal{E} = (\delta u,[\hat{\mathcal{H}}_0 - E]\delta u) - 2D[\delta n,\delta n],
\end{equation}
where the bilinear functional $D$ is defined in Eq.~(\ref{DOperator}) and $\delta n = 2\Re{\chi_0^*\delta u}$. First, from the definition of $\hat{\mathcal{H}}_0$ and using integration by parts we get
\begin{equation}
\label{KineticPart}
   (\delta u,[\hat{\mathcal{H}}_0 - E]\delta u) =  (\nabla \delta u, \nabla \delta u) + (\delta u, [U_0 - E]\delta u),
\end{equation}
which shows that this term is well-defined for any $\delta u\in H^1(\Real^3,\Complex^{2\ell+1})$ lying in the Hilbert space of functions $\delta u: \Real^3\to \Complex^{2\ell+1}$ such that $\delta u$ and $\nabla\delta u$ are quadratically Lebesgue-integrable. Sobolev's inequality \cite{brezis2011functional} implies that the components of $\delta u$ are in $L^p(\Real^3,\Complex)$ for any $2\leq p\leq 6$. Since the same properties hold true for $\chi_0$, it follows that $\delta n\in L^q(\Real^3,\Real)$ for any $1\leq q\leq 3$.

Next, we use Young's convolution inequality \cite{lieb2001analysis} to estimate $D[\delta n, \delta n]$. In order to do this we write it as follows
\begin{equation}
\label{DConvolutionForm}
D[\delta n, \delta n] = \frac{1}{16\pi}\int \delta n(x) (w * \delta n)(x) d^3x, 
\end{equation}
where $*$ refers to the convolution operation and $w(x) := 1/|x|$. Next, decompose $w = w_1 + w_2$ where $w_1(x) := 1/|x|$ for $0 < |x| < R$ and $w_1(x) = 0$ for $|x| \geq R$  with $R > 0$ a free parameter that we will choose later. The functions $w_1$ and $w_2$ have $p$-norms $\| \cdot \|_p$ given by
\begin{equation}
    \|w_1\|_{3/2} = c_0 R, \quad \|w_2\|_{\infty} = \frac{1}{R}, \quad c_0 = \left(\frac{8\pi}{3}\right)^3.
\end{equation}
Therefore, Young's convolution inequality implies that
\begin{equation}
    \label{EstimateD}
   D[\delta n,\delta n] \leq \frac{\|\delta n\|_1}{16\pi} \left( c_0 R \|\delta n\|_3 + \frac{1}{R} \|\delta n\|_1\right).
\end{equation}
Using the Cauchy-Schwarz inequality, the norm $\|\delta n\|_1$ can be estimated as follows:
\begin{align}
    \|\delta n\|_1 &= 2\int \left|\Re\left[\frac{f(\vec{x})\chi_0^*(\vec{x})\delta u(\vec{x})}{f(\vec{x})}\right]\right| d^3x\\ \nonumber
    &\leq 2\int |f(\Vec{x}) \chi_0(\Vec{x})| \left|\frac{\delta u (\Vec{x})}{f(\Vec{x})}\right| d^3x \leq 2\|f \chi_0\|_2 \|\delta u/f\|_2,
\end{align}
where $f$ is an arbitrary positive function such that $f\chi_0$ and $\delta u / f$ are square integrable. In a similar way, one obtains
\begin{equation}
\|\delta n\|_3 \leq 2\|\chi_0\|_6 \|\delta u\|_6 
\leq 2\tilde{C}_1\|\chi_0\|_6     
    \sqrt{(\nabla \delta u, \nabla \delta u)},
\end{equation}
where we have used Sobolev's inequality \cite{brezis2011functional} in the last step with a corresponding positive constant  $\tilde{C}_1 > 0$. Using these estimates in the inequality~(\ref{EstimateD}) yields
\begin{equation}
    D[\delta n, \delta n] \leq \tilde{C}_2 R \sqrt{(\nabla \delta u, \nabla \delta u)} \|\delta u/f\|_2 + \frac{\tilde{C}_3}{R} \|\delta u/f\|_2^2,
\end{equation}
with positive constants $\tilde{C}_2$ and $\tilde{C}_3$ depending on $f$. Combining this result with Eq.~(\ref{KineticPart}) and the well-known inequality $2ab \leq a^2 + b^2$ we obtain the following estimate for the second variation of $\mathcal{E}$:
\begin{align}
    \delta^2 \mathcal{E} &\geq (\delta u, (U_0 - E)\delta u) + (1 - \tilde{C}_2 R) (\nabla \delta u, \nabla \delta u)\\ \nonumber
    &- \left(\tilde{C}_2 R + \frac{2\tilde{C}_3}{R}\right) \|\delta u/f\|_2.
\end{align}
Fixing $R$ in such a way that $\tilde{C}_2 R = 1/2$ and defining $C_1 := \tilde{C}_2 R + 2\tilde{C}_3 / R$ finally implies the desired estimate
$$
\delta^2 \mathcal{E} \geq \frac{1}{2} (\nabla \delta u, \nabla \delta u) + (\delta u, [U_0 - E]\delta u)
 - C_1 \|\delta u/f\|^2_2.
$$

%%%%%%%%%%%%%%%%%%%%%%%%%%%%%%%%%%%%%%%
\section{First-order form of the perturbation equations, regularity at the center, and validation of the numerical code}
\label{App:FirstOrderFormulation}
%%%%%%%%%%%%%%%%%%%%%%%%%%%%%%%%%%%%%%%
\begin{figure*}[tb]
	\centering	
\includegraphics[width=\linewidth]{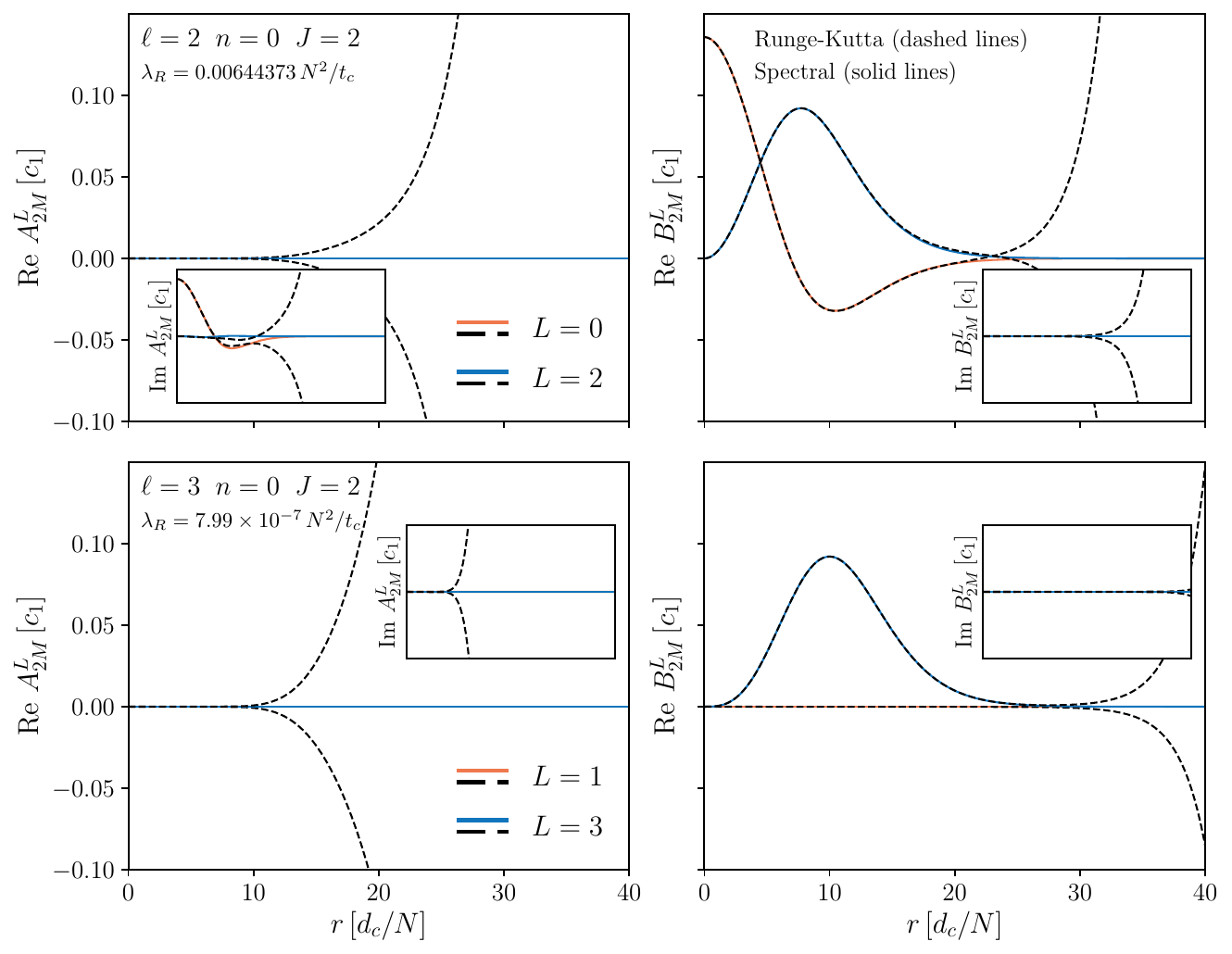}
	\caption{Comparison of the linearized modes $(\mathcal{A}_{JM}, \mathcal{B}_{JM})$ obtained from the spectral (solid lines) and Runge-Kutta (dashed lines) methods for the background configuration with $\ell= 2, 3$ and $n=0$ and two real eigenvalues corresponding to the row $J=2$ in Table~\ref{Fig/Tab2}.}\label{FigRK}
\end{figure*}

In this final appendix, we rewrite the perturbation equations~(\ref{EqGPN}) as a first-order system of ordinary differential equations with a regular singular point at $r=0$. This allows us to prove that the perturbation equations possess solutions satisfying the desired regularity properties near the center. Furthermore, by performing an independent Runge-Kutta integration of it, we use this system to validate the numerical results obtained in section~\ref{Sec:Numerical}. We assume that the pair $JM$ has been fixed, and to alleviate the notation,  we shall omit the corresponding subscripts. Hence, in the following, we write $(a^L,b^L)$ instead of $(a_{JM}{}^L,b_{JM}{}^L)$ etc.

The first-order system is obtained from Eqs.~(\ref{EqGPN}) by introducing the following fields:
\begin{equation}
X^L := r^{-L-1} a^L,\qquad
Y^L := r^{-L-1} b^L,
\end{equation}
and
\begin{equation}
Z^L := 2r^{-J-1} 
\bigg(\frac{d^2}{dr^2}-\frac{J(J+1)}{r^2}\bigg)^{-1}\left[\sigma_\ell^{(0)} a^L \right],
\end{equation}
as well as
\begin{equation}
\xi^L := dX^L/dr,\quad 
\eta^L := dY^L/dr,\quad
\zeta^L := dZ^L/dr.
\end{equation}
This yields the system
\begin{subequations}\label{FirstOrderSyst}
\begin{align}
\frac{d}{dr} X^L &= \xi^L,\\
\frac{d}{dr} Y^L &= \eta^L,\\
\frac{d}{dr} Z^L &= \zeta^L,\\
\frac{d}{dr}\xi^L &= - \frac{2(L+1)}{r}\xi^L - u^{(0)} X^L - i\lambda Y^L + Q^L,
%\nonumber\\
% &+ r^{J-L}\sigma_\ell^{(0)}\sum\limits_{L'=|J-\ell|}^{J+\ell}
%\frac{\sqrt{(2L+1)(2L'+1)}}{2J+1} C^{J0}{}_{L0\ell 0} C^{J0}{}_{L'0\ell 0} Z^{L'},
\\
\frac{d}{dr}\eta^L &= - \frac{2(L+1)}{r}\eta^L - u^{(0)} Y^L 
 - i\lambda X^L,\\
\frac{d}{dr}\zeta^L  &= -\frac{2(J+1)}{r}\zeta^L + 2r^{L-J}\sigma_\ell^{(0)} X^L,
\label{Eq:zetaL}
\end{align}
\label{Eq:FOSystem}
\end{subequations}
where
\begin{align}
Q^L &:= r^{J-L}\sigma_\ell^{(0)}
\nonumber\\
&\times\sum\limits_{L'=|J-\ell|}^{J+\ell}
\frac{\sqrt{(2L+1)(2L'+1)}}{2J+1} C^{J0}{}_{L0\ell 0} C^{J0}{}_{L'0\ell 0} Z^{L'}.
\label{Eq:QLDef}
\end{align}
Note that it is possible to reduce the number of equations by replacing the fields $Z^L$ with the single field
\begin{equation}
\tilde{Z} :=
\sum\limits_{L'=|J-\ell|}^{J+\ell}
\frac{\sqrt{2L'+1}}{2J+1} C^{J0}{}_{L'0\ell 0} Z^{L'},
\end{equation}
and similarly for $\zeta_{JM}{}^L$. 

Since $\sigma_\ell^{(0)}\sim r^\ell$ near the center and $L$ varies between $|J-\ell|$ and $J+\ell$ the two terms $r^{L-J}\sigma_\ell^{(0)}$ and $r^{J-L}\sigma_\ell^{(0)}$ appearing in the right-hand sides of Eqs.~(\ref{Eq:zetaL}) and (\ref{Eq:QLDef}) are regular at $r=0$, and hence it follows that the first-order linear system~(\ref{Eq:FOSystem}) has a regular singular point at $r=0$~\cite{Walter-Book}. In particular, given any real values for $x^L$, $y^L$, and $z^L$ there exists a unique solution of~(\ref{Eq:FOSystem}) such that
\begin{subequations}
\begin{align}
& X^L(0) = x^L,\quad
Y^L(0) = y^L,\quad
Z^L(0) = z^L,\\
& \xi^L(0) = \eta^L(0) = \zeta^L(0) = 0.
\end{align}
\end{subequations}

Next, we numerically integrate the first-order system~(\ref{FirstOrderSyst}) from the origin outwards using the same adaptive Runge-Kutta integration method as the one employed to obtain the background fields $(\sigma_{\ell}^{(0)}, u^{(0)})$. The boundary values  $x^L, y^L$, and $z^L$ are read off from the respective eigenfields $(a^L, b^L)$ associated with the eigenvalue $\lambda$ computed from the spectral method.
Figure~\ref{FigRK} shows a comparison between the results from the Runge-Kutta integration (dashed lines) and the ones from the spectral method (solid lines) for the modes with $J=2$ corresponding to the eigenvalues $\lambda = -0.00644373-5.11 i \times 10^{-13}$ and $7.99\times10^{-7}+1.37i\times 10^{-14}$ associated with the ground state configurations with $\ell=2, 3$ (see Table~\ref{Fig/Tab2}). We see from this figure that the Runge-Kutta solutions correctly reproduce the relevant parts of the spectral profiles up to some given radius after which they start diverging due to their sensitive dependency on the boundary values (i.e., $x^L, y^L$, and $z^L$) and on the value of $\lambda$.

%%%%%%%%%%%%%%%%
%\subsection*{\label{sec:citeref} References}
\bibliographystyle{unsrt}
\bibliography{ref.bib} 
%%%%%%%%%%%%%%%%
 
\end{document}